\documentclass[sigplan,screen]{acmart}

\AtBeginDocument{%
  \providecommand\BibTeX{{%
    \normalfont B\kern-0.5em{\scshape i\kern-0.25em b}\kern-0.8em\TeX}}}

\setcopyright{acmcopyright}
\copyrightyear{2018}
\acmYear{2018}
\acmDOI{10.1145/1122445.1122456}

\acmConference[Woodstock '18]{Woodstock '18: ACM Symposium on Neural
  Gaze Detection}{June 03--05, 2018}{Woodstock, NY}
\acmBooktitle{Woodstock '18: ACM Symposium on Neural Gaze Detection,
  June 03--05, 2018, Woodstock, NY}
\acmPrice{15.00}
\acmISBN{978-1-4503-XXXX-X/18/06}

\usepackage{dblfloatfix}
\usepackage{multirow}
\usepackage{makecell}
\usepackage{enumitem}
\usepackage{caption}
\usepackage{subcaption}
\usepackage{diagbox}
\usepackage{graphicx}
\usepackage{lipsum}

\setcopyright{none}
\renewcommand\footnotetextcopyrightpermission[1]{} 

\usepackage{algorithm}
\usepackage{algorithmic}

\usepackage{pifont}

\begin{document}

\title{$2^{1296}$ Exponentially Complex Quantum Many-Body Simulation via Scalable Deep Learning Method}

\author{Xiao Liang\textsuperscript{1}\textsuperscript{\ding{61}},
Mingfan Li\textsuperscript{1}\textsuperscript{\ding{61}},
Qian Xiao\textsuperscript{1}\textsuperscript{\ding{61}},
Hong An\textsuperscript{1}\textsuperscript{,}\textsuperscript{2}\textsuperscript{*},
Lixin He\textsuperscript{1}\textsuperscript{*},
Xuncheng Zhao\textsuperscript{1},
Junshi Chen\textsuperscript{1}\textsuperscript{,}\textsuperscript{2},
Chao Yang\textsuperscript{3},
Fei Wang\textsuperscript{4},
Hong Qian\textsuperscript{1},
Li Shen\textsuperscript{1},
Dongning Jia\textsuperscript{2},
Yongjian Gu\textsuperscript{5},
Xin Liu\textsuperscript{4},
Zhiqiang Wei\textsuperscript{2},\textsuperscript{5}}
\thanks{*Corresponding authors. \\
	\ding{61}Equal contributions.}

\affiliation{%
	\institution{\textsuperscript{1} University of Science and Technology of China, Hefei, Anhui, China}
	\city{}
	\state{}
	\country{}
}
\email{lxxhlb@ustc.edu.cn, mingfan@mail.ustc.edu.cn, xqbeida@mail.ustc.edu.cn}
\email{han@ustc.edu.cn, helx@ustc.edu.cn, zhaoxc@mail.ustc.edu.cn, cjuns@mail.ustc.edu.cn}
\email{qh1986@mail.ustc.edu.cn, shenliwww@mail.ustc.edu.cn}

\affiliation{%
	\institution{\textsuperscript{2}  Pilot National Laboratory for Marine Science and Technology (Qingdao), China}
	\city{}
	\state{}
	\country{}
}
\email{jiadn@ouc.edu.cn}

\affiliation{%
	\institution{\textsuperscript{3}  School of Mathematical Sciences, Peking University, Beijing, China}
	\city{}
	\state{}
	\country{}
}
\email{chao_yang@pku.edu.cn}

\affiliation{%
	\institution{\textsuperscript{4}  National Supercomputing Center in Wuxi, Wuxi, China}
	\city{}
	\state{}
	\country{}
}
\email{weedyblues@126.com, yyylx@263.net}

\affiliation{%
	\institution{\textsuperscript{5}  Ocean University of China, Qingdao, China}
	\city{}
	\state{}
	\country{}
}
\email{guyj@ouc.edu.cn, weizhiqiang@ouc.edu.cn}

\begin{abstract}
For decades, people are developing efficient numerical methods for solving the challenging quantum many-body problem, whose Hilbert space grows exponentially with the size of the problem. However, this journey is far from over, as previous methods all have serious limitations. The recently developed deep learning methods provide a very promising new route to solve the long-standing  quantum many-body problems. We report that a deep learning based simulation protocol can achieve the  solution with state-of-the-art precision in the Hilbert space as large as $2^{1296}$ for spin system and $3^{144}$ for fermion system , using a HPC-AI hybrid framework on the new Sunway supercomputer. With highly scalability up to 40 million heterogeneous cores, our applications have measured 94\% weak scaling efficiency and 72\% strong scaling efficiency. The accomplishment of this work opens the door to simulate spin models and Fermion models on unprecedented lattice size with extreme high precision.
\end{abstract}



\maketitle
\pagestyle{plain} 
\fancyhf{}

\section{JUSTIFICATION FOR THE GORDON BELL PRIZE}
Deep-learning based quantum many-body simulations with state-of-the-art lattice size and energy precision, using nearly 40 million heterogeneous cores with 94\% weak and 72\% strong scaling efficiency.
The Hilbert space is as unprecedentedly large as $2^{1296}$ and $3^{144}$, for the spin systems (e.g., $J1$-$J2$ model) and the Fermion systems (e.g., $t$-$J$ model), respectively.
\section{PERFORMANCE ATTRIBUTES}



	\begin{table}[h]
	\resizebox{8cm}{!}{
		\begin{tabular}{cc}
			\hline
			\multicolumn{1}{c}{Performance Attributes}    & \multicolumn{1}{c}{Content} \\ \hline
			\multicolumn{1}{c}{Category of achievement}   &  \multicolumn{1}{c}{Scalability} \\ 		
			\multicolumn{1}{c}{Type of method used}       & \multicolumn{1}{c}{Convolutional neural network}        \\
			\multicolumn{1}{c}{Results reported on basis} & \multicolumn{1}{c}{Whole application}        \\
			\multicolumn{1}{c}{Precision reported}        & \multicolumn{1}{c}{Double precision}        \\
			\multicolumn{1}{c}{Maximum problem Size}      & \multicolumn{1}{c}{$36\times36$ square lattice}        \\
			\multicolumn{1}{c}{System scale}              & \multicolumn{1}{c}{Measured on full system}        \\
			\multicolumn{1}{c}{Measurement mechanism}     & \multicolumn{1}{c}{Timers}        \\ \hline
	
		\end{tabular}
	}
	
\end{table}

\section{OVERVIEW OF THE PROBLEM}
\subsection{The quantum many-body problem}
\begin{figure*}[htbp]
  \centering
  \includegraphics[width=0.9\linewidth]{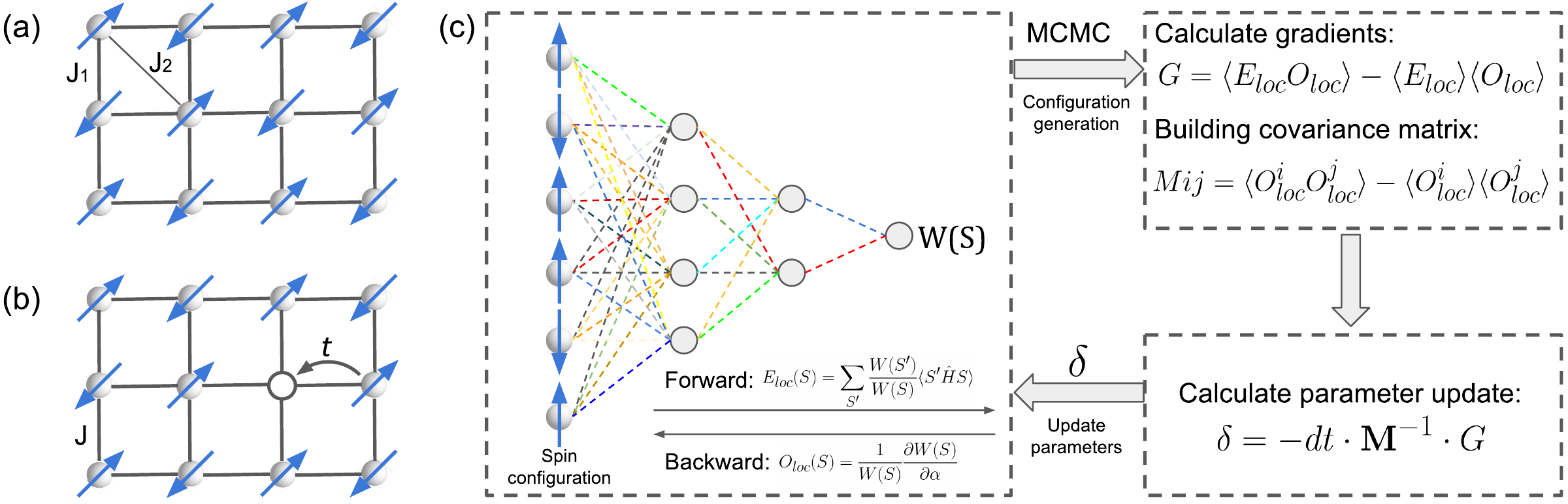}
  \caption{Schematic diagram for $J1$-$J2$ model(a) and $t$-$J$ model(b). (c)The self-learning optimization procedure of the neural network.}
  \label{flowchart}
\end{figure*}

Strongly correlated quantum many-body physics is one of the
most fascinating research fields in condensed matter physics.
When a large number of microscopic particles interact with each other,
quantum entanglement is built among them and exotic physical phenomena emerge,
such as (high-temperature) superconductivity\cite{zhou2021high} reported by Nobel laureates for several times\cite{superconductivity_url}, quantum Hall effects\cite{wen1992theory} reported by Nobel laureates in the year of 1998\cite{quantumhalleffect_url}, quantum spin-liquid\cite{QSL} reviewed in the Science journal in the year of 2020, etc. Solving these problems will greatly deepen our understanding of the
basic laws of the natural world and guide us to find novel physical phenomena
and new quantum materials, which may have great potential applications in energy, information, etc.

Despite the basic rules of quantum mechanics are known, solving the strongly correlated
quantum many-particle problems is still extremely challenging.
This is because the Hilbert space of the solution grows exponentially with the size of the problem.
Furthermore, the perturbative methods, although have made great successes for
simulating weakly correlated material systems,
will completely fail for the strongly correlated systems.
Revealing the fascinating physical natures in quantum many-body systems
mainly relays on the non-perturbative numerical methods,
such as exact diagonalization (ED),
quantum Monte Carlo method (QMC) \cite{Anderson2007}  and density matrix renormalization group (DMRG) \cite{white92}.
However, these methods all have serious limitations: e.g., the computational cost of ED grows exponentially with the system size,
and therefore the system size of the ED method is limited to less than 50 sites\cite{Savary2017}.
QMC suffers from the notorious sign problem for fermionic and frustrated systems\cite{Troyer05};
and DMRG is limited to 1D or quasi-1D systems and does not work well for higher dimension systems\cite{Schollw11}.
To keep the same accuracy, the computational costs of DMRG grow exponentially with the width of the lattice.
So far, the width of DMRG simulation is limited to $L\approx$12\cite{dmrgj1j2}.
Very recently, the so-called tensor network methods, e.g., the projected entangled-pair states (PEPS) method\cite{PEPS} have been developed, which can simulate the fermionic and frustrated systems. Some of the authors in this paper have developed large scale PEPS code PEPS++, and implemented it on Sunway TaihuLight\cite{he2018peps++}. We have demonstrated solving the $J1$-$J2$ model on a $24\times 24$ lattice with open boundary conditions (OBC), with the bond dimension $D=16$. However, the periodic boundary conditions (PBC) usually has lower boundary effects than OBC, there are more suitable for simulating quantum many-body systems. However due to the high scaling to the bond dimension $O(D^{18})$, there is no report of PEPS simulation on PBC with $D\ge 6$, which is still too small to catch the essential physics of quantum many-body models.

\subsection{Deep learning methods for quantum many-body problems}
\begin{table*}
\centering
\caption{Comparisons of current state-of-art neural network methods for solving spin models and the Fermion models. The frustrated-free Heisenberg model has been solved by RBM to a rather high precision\cite{carleo2017solving}. For the $J1$-$J2$ model, three kinds of methods are used: 1, a single CNN\cite{liang2018solving}; 2, a deep CNN\cite{choo2019two,szabo2020neural,liang2021hybrid,li2022bridging}; 3, a Gutzwiller projected wave-function and a RBM\cite{ferrari2019neural,nomura2021dirac}. Because of the high computational complexity of Gutzwiller projected wave-function, the investigated lattice size is limited to $L$=18\cite{nomura2021dirac}. The method of a deep CNN with has lower computational complexity, however the energy's precision depends on the representation ability of the deep CNN\cite{szabo2020neural,choo2019two,liang2021hybrid}. With a proper deep CNN structure, and increasing the parameter number to the magnitude of $10^5$, the energy's precision reaches state-of-the-art\cite{li2022bridging}. In this work, the parameter number is further increased to the magnitude of $4\times 10^5$, and the investigated lattice size is increased to $L$=36. For the Fermion models, because of the high computational complexity of Slater determinants, the investigated lattice size is limited\cite{luo2019backflow,stokes2020phases,moreno2021fermionic}.
In this work, the $t$-$J$ model is solved on the square lattice with size up to $L$=12, using the determinant free deep CNN wave-functions.}
\label{methods_comparisons}
\begin{tabular}{|c|c|c|c|c|c|}
\hline
Model Type               & \begin{tabular}[c]{@{}c@{}}Quantum\\Model\end{tabular}                  & Method                                                                                             & Year & \begin{tabular}[c]{@{}c@{}}Neural Network\\ Parameter Number\end{tabular} & \begin{tabular}[c]{@{}c@{}}Largest\\ Lattice Size\end{tabular}  \\
\hline
\multirow{7}{*}{Spin}    & Heisenberg                                                              & RBM     \cite{carleo2017solving}                                                                                           & 2017 & 323200                                                                    & 10x10                                                           \\
\cline{2-6}
                         & \multirow{6}{*}{$J1$-$J2$}                                              & CNN\cite{liang2018solving}                                                                                                & 2018 & 11009                                                                     & 10x10                                                           \\
\cline{3-6}
                         &                                                                         & Gutzwiller Projection and RBM\cite{ferrari2019neural}                                                                      & 2019 & 2000                                                                      & 10x10                                                           \\
\cline{3-6}
                         &                                                                         & Sign Rule and CNN \cite{choo2019two}                                                                                 & 2019 & 7676                                                                      & 10x10                                                           \\
\cline{3-6}
                         &                                                                         & Pair Product State and RBM\cite{nomura2021dirac}                                                                         & 2021 & 5200                                                                      & 18x18                                                           \\
\cline{3-6}
                         &                                                                         & Sign Rule and CNN\cite{li2022bridging}                                                                                 & 2022 & 106529                                                                    & 24x24                                                           \\
\cline{3-6}
                         &                                                                         & Sign Rule and CNN (\textbf{This Work})                                                             & 2022 & 421953                                                                    & 36x36                                                           \\
\hline
\multirow{5}{*}{Fermion} & \begin{tabular}[c]{@{}c@{}}Spinless\\Fermion\end{tabular}               & \begin{tabular}[c]{@{}c@{}}Slater Determinants, CNN Jastrow\cite{stokes2020phases}\\ and CNN sign Correction\end{tabular} & 2020 & -                                                                         & 10x10                                                           \\
\cline{2-6}
                         & \multirow{3}{*}{\begin{tabular}[c]{@{}c@{}}Fermi\\Hubbard\end{tabular}} & Slater Determinants with FNN\cite{luo2019backflow}                                                                       & 2019 & $\approx$4880                                                             & 4x4                                                             \\
\cline{3-6}
                         &                                                                         & Determinants Free FCNN\cite{inui2021determinant}                                                                             & 2021 & 2070120                                                                   & 6x6                                                             \\
\cline{3-6}
                         &                                                                         & Slater Determinants with FNN\cite{moreno2021fermionic}                                                                       & 2022 & -                                                                         & 4x16                                                            \\
\cline{2-6}
                         & $t$-$J$                                                                 & Determinants Free CNN (\textbf{This Work})                                                         & 2022 & 113815                                                                    & 12x12                                                           \\
\hline
\end{tabular}
\end{table*}

Recently, the rapid progress of artificial intelligence powered by deep learning\cite{silver2016mastering,silver2017mastering} has attracted science community on applying deep learning to solve scientific problems. For example, using neural network to solve differential equations\cite{doi:10.1073/pnas.1718942115}, accelerate molecular dynamics\cite{jia2020pushing}, predict proteins' 3D structures\cite{jumper2021highly}, control nuclear fusion\cite{degrave2022magnetic}, and so on. There are also great efforts in applying
the deep learning methods to study the quantum many-body problems.


Take the quantum spin model as an example, a quantum many-body state is represented by a superposition of the spin configurations,
\begin{equation}
|\Psi\rangle=\sum_S W(S)|S\rangle,
\label{psi}
\end{equation}
where $|S\rangle=|s_1^z,s_2^z,\cdots,s_N^z\rangle$ is a set of spins with one on each site of the lattice, and $N$ is the total number of sites. If the degree of freedom on each site is $k$, the Hilbert space (i.e., the number of $|S\rangle $) of the problem is $k^N$.  The weight of the spin configuration $W(S)$ is represented by a neural network. The network parameters are optimized to determine the ground state of the system, i.e., the state of the lowest energy $E_{\rm g.s}={\rm min}\langle\Psi\hat{H}\Psi\rangle/\langle\Psi|\Psi\rangle$, where  $\hat{H}$ is the Hamiltonian of the system.

The self-learning procedure for optimizing the neural network is depicted in Fig.(\ref{flowchart})(c). Firstly, Markov-Chain-Monte-Carlo (MCMC) is performed according to $|W(S)|^2$. When collecting a sample, the forward of neural network gives local energy $E_{loc}$ and the backward of neural network gives local derivative $O_{loc}$. Secondly, the gradients and the covariance matrix needed for the so-called Stochastic Reconfiguration (SR) optimization method\cite{SR} are calculated. Finally, the network parameters are updated, and the values for updating parameters are $\delta$. Within the fast computation speed of neural networks, the wave-functions can be further optimized by a Lanczos step\cite{hu2013direct}.

Comparing to the traditional deep learning tasks such classifications, there are some major challenges to solve the quantum many-body problems via neural networks,
because one has to obtain an extremely high accurate ground state in the exponentially large Hilbert space:
\begin{itemize}
\item The neural network's generalization ability should be high enough to represent the quantum state in the exponentially large Hilbert space.
\item Double precision of the network parameters is mandatory.
\item The ground energy should be the global minimal of the energies. The first-order-gradient based optimizers like Adam, SGD, et.al are not efficient, as they can be easily trapped into local minimal. Here, a second-order natural-gradient-like method, such as the Stochastic Reconfiguration method is used.
\item Finding the solution in the exponentially-large Hilbert space requires extremely large amount of MCMC samples, which is also required for the precise SR optimization.
\end{itemize}

\section{CURRENT STATE OF THE ART}

Great efforts have been made to  solve the quantum many-body problems via deep learning method\cite{ML_physics_review1,ML_physics_review2}.
The comparisons of current state-of-the-art results are listed in Tab.\ref{methods_comparisons}.
In the year of 2017, Carleo et al. solved the Heisenberg model via Restricted-Boltzmann-Machine (RBM), with 323200 network parameters and obtained rather high precision results on the $10\times 10$ square lattice\cite{carleo2017solving}. However, the NN can not handle frustrated systems, such as the famous $J1$-$J2$ model, which is depicted in Fig.(\ref{flowchart})(a).
Liang and He \cite{liang2018solving} solved the frustrated $J1$-$J2$ model using a single-layered CNN with 11009 parameters with satisfactory precision.
To further improve the energy precision on the $10\times 10$ lattice, the Gutzwiller Projection fermionic wave-function with a RBM\cite{ferrari2019neural} and the deep CNN with a sign rule\cite{choo2019two} are used. In the year of 2021, on the ``Fugaku" supercomputer, the pair product states with a RBM has achieved high accuracy on the lattice as large as $18\times 18$\cite{nomura2021dirac}. In the year of 2022, by increasing the deep CNN's parameter number to 106529, the energy's precision is highly improved and the lattice size is increased to $24\times 24$\cite{li2022bridging}. In this work, the deep CNN's parameter number is further increased to 421953, and the square lattice size is increased to $L$=36.

The Fermion models are even more challenging. In the year of 2019, by employing a CNN based Jastrow and sign corrections, the spinless Fermions model is investigated on a $10\times 10$ lattice\cite{stokes2020phases}. There are several investigations for the Fermi-Hubbard model. In 2019, by employing a fully-connected neural network (FNN) in the Slater determinant, the model is studied on a $4\times 4$ lattice\cite{luo2019backflow}. In the year of 2021, a FCNN  is used as the determinant free wave-function to study
the model on a $6\times 6$ lattice\cite{inui2021determinant}. In the year of 2022, by employing the FNN in the Slater determinants, the lattice size is increased to $4\times 16$\cite{moreno2021fermionic}.
In this work, by using an efficient CNN with the network parameters of 113815, we investigate the $t$-$J$ model, up to a $12\times 12$ lattice.
The $t$-$J$ model is schematically depicted in Fig.(\ref{flowchart})(b).

\begin{figure*}[htbp]
	\centerline{\includegraphics[scale=0.6]{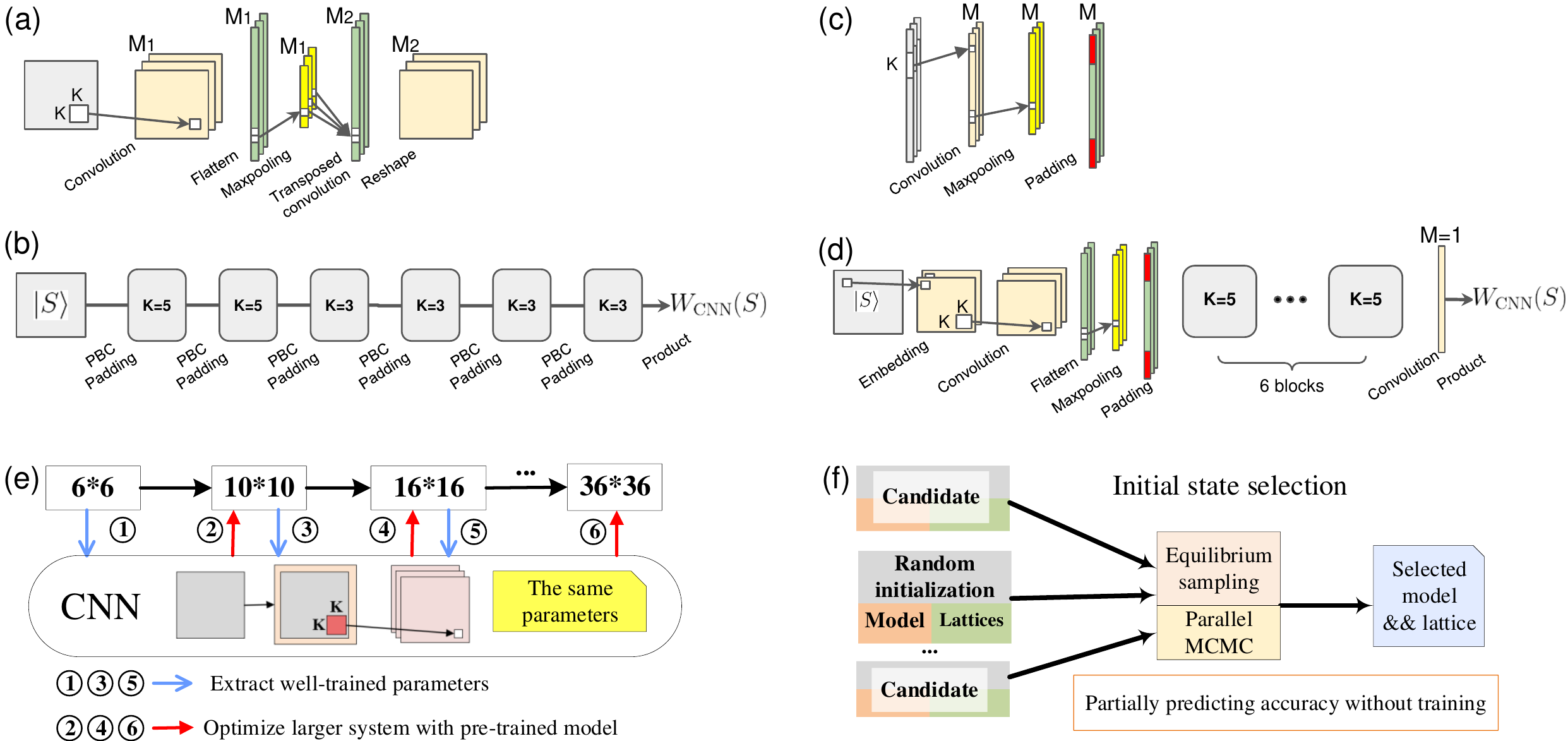}}
	\caption{The structure of CNN1 is depicted by (a)(b), where (a) is the building block the deep structure of (b). The structure of CNN2 is depicted by (c)(d), where (c) is the building block of the deep structure of (d). (e) The transfer learning process of CNN1. (f) The initial state selection process of CNN2.}
	\label{innov-1}
\end{figure*}

\section{INNOVATIONS}
\subsection{Summary of contributions}
We develop a highly efficient and highly scalable method for solving quantum many-body problems to the extremely high precision. This is achieved by combining the unprecedented representation ability of the deep CNN neural network structures (Fig.\ref{innov-1}(a)-(d)), and a highly scalable(Fig.\ref{innov-1}(e)(f)) and fine-tuned implementation on heterogeneous Sunway architectures(Fig.\ref{innov-2}). With the optimized HPC-AI framework and up to 40 million heterogeneous sw26010pro cores, we solve the ground state of the  $J1$-$J2$ model in the 36$\times$36 lattice, corresponding to a 2$^{1296}$ dimensional Hilbert space, and solve the $t$-$J$ model in the $12\times 12$ lattice, corresponding to a $3^{144}$ dimensional Hilbert space.

\subsection{Neural network innovations}
To meet the strictly demanding requirement, we develop two deep CNN structures for the spin models and fermion models, respectively.  We denote the two NN structures, CNN1 and CNN2 respectively. The computational complexity of CNN is much lower comparing to the fully connected structure. By taking the advantage of translational invariance, the CNN can scale to very large lattices.

The structure of CNN1 is depicted in Fig.(\ref{innov-1})(a)(b), the network is built by stacking the building block denoted in Fig.(\ref{innov-1})(a) for six times\cite{liang2021hybrid}. A building block consists the two-dimensional convolution and one-dimensional maxpooling and transposed-convolution. To maintain the dimensions, paddings are employed on the input spin lattice and between two building blocks. The padding scheme is based on the periodic boundary conditions. The final output is the product of the neurons, which is the wave-function coefficient for the input spin configuration.

The structure of CNN2 is depicted in Fig.(\ref{innov-1})(c)(d). CNN2 is originated from CNN1, with two major modifications: 1, the input spin lattice is processed by an embedding layer; 2, the one-dimensional maxpooling is performed with stride one.
To maintain the dimensions, padding is performed by copying the first several neurons to the last. The deep structure is built by stacking the building block in Fig.(\ref{innov-1})(c) for six times. Finally, with a convolution operation, the channel number is decreased to one, the final output is the product of the neurons.

The deep CNN structures used in this work have several important differences compared to other CNN structures for quantum many-body problems.\cite{carleo2017solving,choo2019two,stokes2020phases}
First, the nonlinearity in our deep CNN is induced by the maxpooling, instead of traditional activation functions. The maxpooling picks up the most important
degree of freedom in a convolution filter, which is similar to the coarse-grained process in a renormalization group theory.\cite{DMRG}. Secondly, the wave-function coefficient is generated by the product of neurons, which differs from the exponential function in RBM based structures.

\subsection{Algorithm innovations}

The training process of deep learning usually start from random parameters. A good initial state may greatly accelerate the training process, especially for
the quantum many-body problems, which have exponentially-large Hilbert space. A good initial state is also benefit for avoiding local optimizations and
therefore drastically reduces the computational cost, especially for large-scale lattices. We develop several techniques to select the initial states.

\subsubsection{Transfer learning for large lattice}
Because of the exponentially large Hilbert space, direct training of the model in large lattice starting from random parameters is very difficult.
The convolution operation is intrinsically scalable for different lattice sizes.
Furthermore, the energy convergence on a small lattice is fast, as the lattice size increases, the ground energy will converge with respect to the lattice size.
Fig. \ref{innov-1}(e) presents transfer learning technique applied in this work. Initially, the network is trained on $6\times 6$ lattice from randomly initialized parameters. The network parameters are then checkpointed and used for optimizing large $10\times 10$ system as the pre-trained model. With multiple stages of transfer learning, we finally obtain a good initial state for solving the ground state on $36\times 36$ lattice. Thanks to transfer learning, the local optimizations are avoided and the training steps for large lattice are significantly reduced.

\subsubsection{Parallel initial state selection}
Because of the exponentially large Hilbert space, a proper initial model parameter and the initial spin configuration is beneficial for avoiding local optimizations, thus crucial for fast energy convergence.
 Here we calculate the energy of a randomly initialized CNN by performing MCMC sampling, assuming that the CNN with lower initial energy will have a better convergence after SR optimizations.
The initial state selection for $t$-$J$ model is shown in Fig. \ref{innov-1}(f). To adapt to the heterogeneous architecture, the paradigm can be described in two-level: the different processes have different initial CNN parameters and the independent chains in each process have different initial spin lattice. After the MCMC process, the CNN parameter together with the initial spin configuration that has the lowest energy result are selected. After the initial state selections, the optimization steps are significantly reduced especially on large lattices.

\subsection{HPC-AI hybrid framework innovation}
Despite the fact that quantum many-body problem resides in the realm of dimensional reduction and feature extraction in a much border context, current deep learning frameworks and model optimizations are inefficient. There are two major difficulties: Firstly, the input size and the network size applied in physics are usually smaller than that of mainstream deep learning tasks, where insufficient computation leads to low computing efficiency and hardware utilization. Secondly, the global minimal energy is difficult for the prevalent first-order gradient optimizer (i.e., Adam, SGD) and the mini-batch based training. To address above difficulties, we propose a highly optimized HPC-AI hybrid framework on the new Sunway supercomputer and achieve multifaceted improvements.

\subsubsection{MCMC-SR parallel framework}
The whole application can be divided into two components: the front-end MCMC sampling and the back-end model optimization. The first stage can be seen as data parallelism, where different processes hold the same model parameters but compute different input data. The second stage is responsible to handle MCMC samples and compute the $\delta$ of network parameters in SR by constructing a covariance matrix. 

\begin{figure*}[htbp]
	\centerline{\includegraphics[scale=0.37
	]{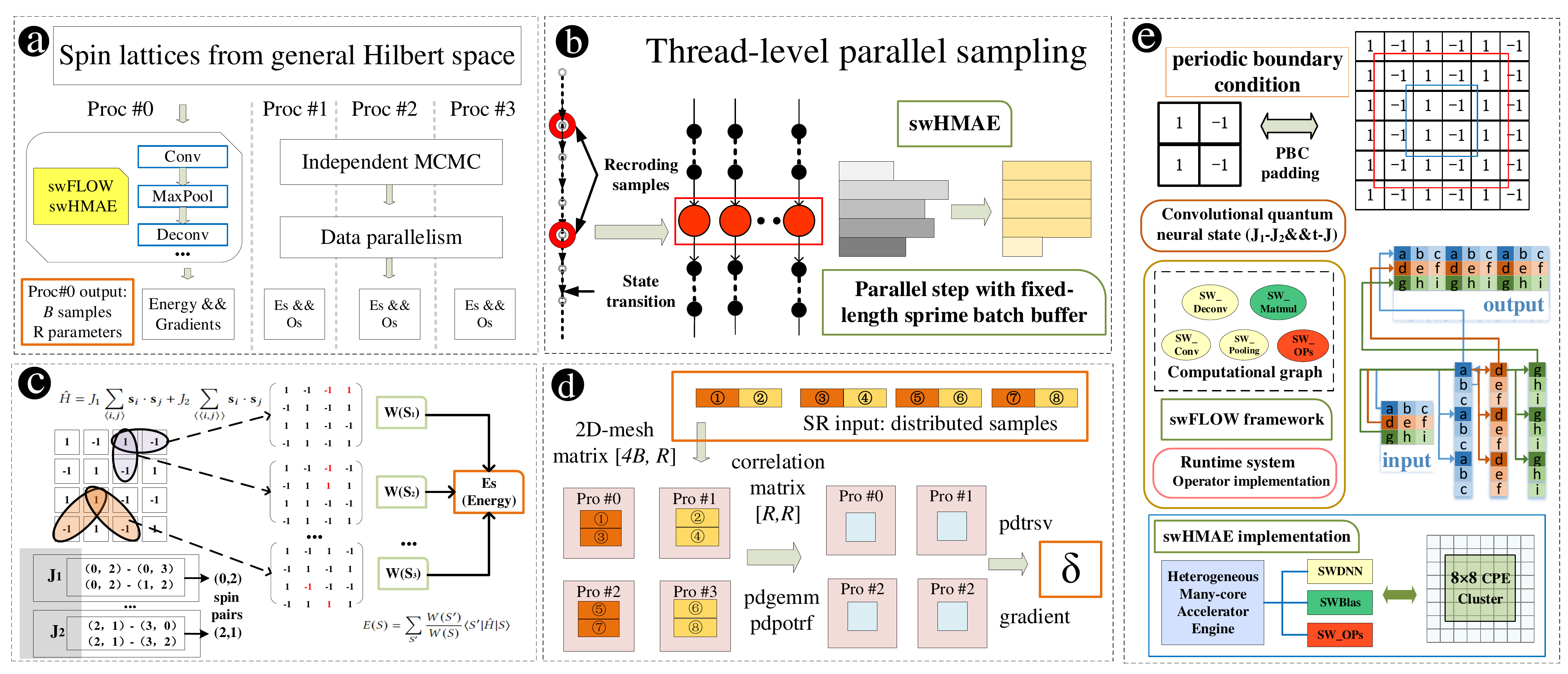}}
	\caption{Overview of the HPC-AI hybrid framework. (a) Data parallelism for MCMC sampling. (b) Batch-enhanced CNN execution. (c) Local energy calculation for $J1$-$J2$ model. (d) Flowchart of 2D-mesh distributed SR. (e) Optimized PBC padding operator for swHMAE and swFLOW.}
	\label{innov-2}
\end{figure*}

\textbf{Process-level parallel MCMC}
Fig. \ref{innov-2}(a) depicts the parallel MCMC sampling among multiple processes. The importance sampling in the Hilbert space is achieved by multiple independent Markov chains. Considering the MC sampling tasks are intrinsically parallel, they can be distributed across all participated processes. Inside each process, the model forward and backward generate $E_{loc}$ and $O_{loc}$, respectively. The computation of the deep CNN is divided layer by layer in the deep learning framework swFLOW. On sw26010pro, the major computation on MPE is accelerated by the swHMAE with 64 CPEs. The detailed discussion on swFLOW and swHMAE will be presented in
Sec. \ref{swFLOW-swHMAE} . The distributed MCMC sampling of four processes will generate $4\times B$ samples, when each process owns $B$ samples. Each sample contains $R$ parameters, where $R$ is the number of network parameters.

\textbf{Thread-level parallel MCMC}
By increasing batch size in the model forward and backward process, the hardware utilization can be further increased. One process maintains multiple independent chains and obtains multiple sampling states in one step, which greatly increases the independent chain number and improves the overall performance. Meanwhile, multiple repeated CNN forward with varying lengths can be replaced by the fixed-length execution, seen in Fig. \ref{innov-2}(b). With the same batch size, the repeated model forward executions may reuse the existing tensor memory resource and improve the performance with efficient static graph execution mode. Such implementation increases the sample number and leads to faster energy convergence.

\textbf{Hotspot energy calculation}
The most time-consuming step in each iteration is to collect the local energy $E_{loc}$, which is about $O(L^2)$ complexity for $L\times L$ lattice. Specifically, obtaining the correct energy of each lattice requires tremendous computation for sprime\_batch ($\sim L^2$ for $E_{loc}$) and sampling gap ($\sim L^2$ for random walking). Therefore the model forward execution is roughly $4L^2$ times that of backward, which differs from mainstream tasks (each forward execution corresponds to one backward). For the $J1$-$J2$ model with $L$=36 square lattice, one optimization step with millions of independent samples requires billions of CNN forward execution, thus the calculation of CNN forward comprises over 95\% of total execution time.

Taking the spin-1/2 $J1$-$J2$ model for instance, the lattice sampling in Fig.\ref{innov-2}(c) presents an example of $4\times 4$ lattice. The spin lattice is wrapped as a 2D tensor (the input image) for model input. Each pixel in the image represents the spin value on the corresponding lattice site, where the value of each pixel is $s_i=\pm 1$. The whole image represents a spin configuration, which is a basis of the wave-function: $|s_1,s_2,\cdots,s_{16}\rangle$. The local energy is closely related to Hamiltonian. The process for calculating a local energy $E_{loc}$ is denoted in Fig.\ref{innov-2}(c). For example, considering the nearest-neighbor interactions, when the spin pair between nearest-neighbor sites ($\langle i, j\rangle$) owns different spin values ([-1, 1] for $J1$-$J2$ model or [-1, 0, 1] for $t$-$J$ model), a new spin configuration with flipping the spin pairs is collected. In the figure, three possible spin configurations according to the proposed spin pairs are marked in red. Considering all the terms in the Hamiltonian, the batch size of collected spin configurations is in the magnitude of $O(L^2)$. These spin configurations are then fed into the CNN model and then generate the logits ($W(S)$). The local energy is calculated according to $E_{loc}(S)=\sum_{S'}\frac{W(S')}{W(S)}\langle S'\hat{H}S\rangle$.




\textbf{Distributed SR computation}
With sufficient samples from parallel Markov chains, the values for updating network parameters $\delta$ are calculated. In Fig.\ref{innov-2}(d), the procedure for computing $\delta$ includes three steps: adjusting the data format, constructing the covariance matrix, and solving a system of linear equations. After MCMC sampling, in each process there are a series of samples, and the samples are then organized as a 2D-mesh distributed style. For the system with $P$ processes, the collected samples are divided into $\sqrt{P}$ blocks, and the processes are split into $\sqrt{P}$ groups, where each group contains $\sqrt{P}$ processes. The data block is scattered into corresponding processes within the group. For the instance of four processes (Proc-0 and Proc-1 in group-1, Proc-2 and Proc-3 in group-2), the block exchange between 2 processes can be illustrated by swapping blocks of \ding{173} and \ding{174} in group 1 (\ding{177} and \ding{178} in group 2).
With 2D-mesh distributed matrix [$4B$, $R$], the following equations construct 2D-mesh matrix [$R$, $R$] with parallel matrix multiplication and inversion. The final $\delta$ is obtained by combining the matrix with energy's first-order gradient. Similar to data parallelism training, all processes shares the same $\delta$ by MPI broadcast.

\subsubsection{Deep learning software stack}
\label{swFLOW-swHMAE}
Fig.\ref{innov-2}(e) presents the overall software stack of swFLOW, which is a deep learning framework to support the efficient running of AI applications on Sunway platform.
With standard computational graph design, various neural network models can be easily realized. For adapting swFLOW to the sw26010pro processors, the neural network operator executions are accelerated by Sunway heterogeneous many-core accelerator engine(swHMAE). For typical operators such as convolution, matrix multiplication, they are supported by libraries such as swDNN and swBLAS. In addition, the customized many-core algorithm optimization is directly implemented in swHAME, such as Tile and Slice or other customized exclusive OPs.

Besides the common operators in deep learning, in the CNNs used in this work, some operators like the PBC padding needs to be specially treated. Where the padding scheme is depicted in Fig.\ref{innov-2}(e), the $2\times2$ tensor is padded into the $4\times 4$ tensor (marked in red). The original tensor is tiled on height and width dimensions, and we then extract the central part of the extended tensor. As depicted in Fig.\ref{innov-2}(e), the input tensor is firstly split into many sub-tensors, where each sub-tensor has three elements, such as [a,b,c]. Then we use DMA with broadcast to copy each sub-tensor from the main memory to LDM space. Finally, each CPE will use DMA to move LDM data to the main memory corresponding to the output tensor. In this way, the memory access bandwidth and parallelism of heterogeneous cores can be fully utilized.



%

\section{PERFORMANCE MEASUREMENTS}
\subsection{Physical systems}
In this work, we benchmark our methods
on two important and  representative models, namely the $J1$-$J2$ frustrated spin model [Fig.\ref{flowchart}(a)] and the $t$-$J$ fermion model [Fig.\ref{flowchart}(b)].
The $J1$-$J2$ model is a candidate model for the quantum spin liquids\cite{LIU2022,nomura2021dirac}, whereas the $t$-$J$ is the basic model for the high-temperature superconductivity.
In this work, both models are investigated on the $L$$\times$$L$ square lattices with periodic boundary conditions.

The Hamiltonian of $J1$-$J2$ model reads:
\begin{equation}
	\hat{H}=J1\sum_{\langle i,j\rangle}\textbf{s}_i\cdot\textbf{s}_j+J2\sum_{\langle\langle i,j\rangle\rangle}\textbf{s}_i\cdot\textbf{s}_j
	\label{J1J2}
\end{equation}
where $\langle i,j \rangle$ and $\langle\langle i,j \rangle\rangle$ indicate the nearest and next-nearest neighbouring spins pairs. $\textbf{s}_i$ is the spin operator on the $i$-th site. We set $J1$=1 and $J2=0.5$ throughout the investigations.

The Hamiltonian of $t$-$J$ model is:
\begin{equation}
	\hat{H}=-t\sum_{\langle i,j\rangle,\sigma}(c_{i,\sigma}^\dag c_{j,\sigma}+h.c)+J\sum_{\langle i,j\rangle}(\textbf{s}_i\cdot\textbf{s}_j-\frac{n_in_j}{4})
	\label{tJ}
\end{equation}
where $\langle i,j \rangle$ indicate the nearest neighbouring pairs. The operator $c_{i,s}^\dag$ creates an electron of spin $\textbf{s}$ on site $i$, $\textbf{s}_i$ and $n_i$ are the electron spin moment and charge density operators on site $i$. We encode the Fermions into spins via Jordan-Wigner transformation \cite{JW}. In the simulations, $t$=1, $J$=0.4 and the hole doping $n_h$=0.125 is used.

\subsection{HPC System and Environment}

\begin{figure}[bhtp]
	\centerline{\includegraphics[scale=0.31]{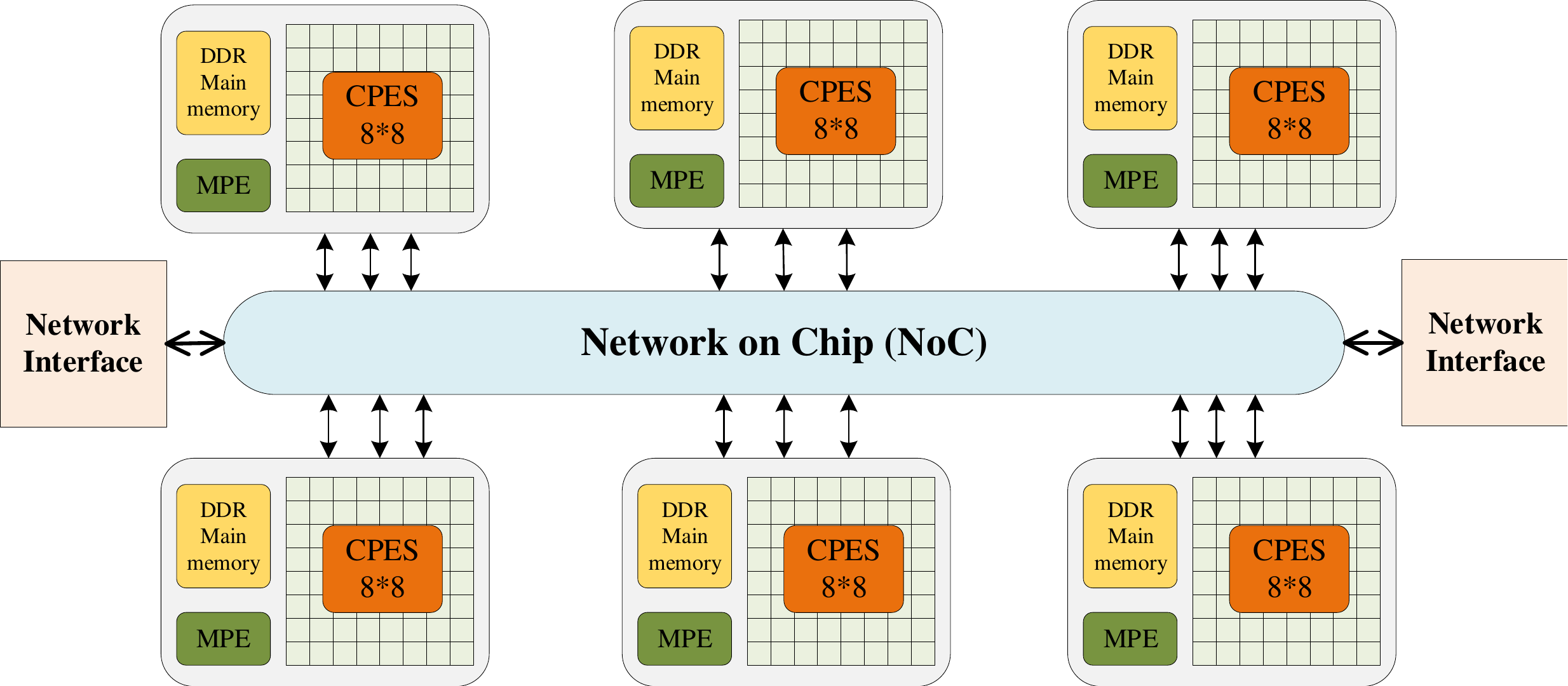}}
	\caption{Architecture of sw26010pro many-core processor.}
	\label{sw26010pro}
\end{figure}

The sw26010pro processor is the upgraded version of the sw26010 processor for the Sunway Taihulight supercomputer.
Fig.\ref{sw26010pro} presents the architecture of sw26010pro, which consists of six core groups (CG), where each CG is attached to a ring network. Each CG contains one management process element (MPE, control core) and a cluster of 64 computing process elements (CPE, computing core) arranged in $8\times 8$ mesh.

Generally, the MPE is responsible for the management of all computing resources and global memory, while the CPE is responsible for executing the tasks assigned by MPE. Besides, a local directive memory (LDM) is attached to each CPE, similar to the shared memory in GPU.

The software environment has been greatly improved on the new Sunway supercomputer, especially for AI applications\cite{fang2017swdnn,li2021swflow,lin2019swflow}. Meanwhile, the traditional high performance algebra libraries, like BLAS, Lapack, and the distributed ScaLapack\cite{choi1992scalapack}, are also optimized. The new generation of Sunway supercomputer delivers great potential to attack various challenges for both HPC and AI.

\subsection{Measurement Methodology}
To achieve high-accuracy ground state energy, double precision floating point numbers are required for faithful quantum system simulations. Since the overall software stack is very complicated, varying from deep learning (swFLOW, swHMAE and swDNN) to classic HPC environment (swBLAS, swLAPACK, swScaLAPACK), manually counting all double-precision arithmetic instructions is impractical. Instead, we collect the number of double-precision arithmetic operations by using the hardware performance monitor provided by the vendor of new Sunway supercomputer.

\section{PERFORMANCE RESULTS}

\begin{table}[thbp]
	\caption{Experimental configuration.}
	\resizebox{8.3cm}{!}{
		\begin{tabular}{|c|c|c|}
			\hline
			\multirow{3}{*}{Application}     & quantum many-body model          & $J1$-$J2$,$t$-$J$                                             \\ \cline{2-3}
			& CNN parameter number & 16443,106529,421953
			\\ \cline{2-3}
			& 2D spin lattice sizes       & $8\times8$ $\sim$ $36\times36$
			\\ \hline
			\multirow{3}{*}{\begin{tabular}[c]{@{}c@{}}System\\environment\end{tabular}}
			& Overall software stack                & BLAS,swDNN,ScaLapack
			\\ \cline{2-3}
			& Maximum number of CGs                  & over 600000
			\\ \cline{2-3}
			& Maximum number of cores                & nearly 40 million
			\\ \hline
		\end{tabular}
	}
	\label{experimental configuration}

\end{table}

\begin{table*}[htbp]
	\centering
	\caption{Scalability results for the MCMC and SR processes.}
	\resizebox{16cm}{!}{
		\begin{tabular}{|c|c|c|c|c|c|c|c|c|c|}
			\hline
			\multirow{2}{*}{model scale} & \multirow{2}{*}{batch size}  & \multicolumn{4}{c|}{MCMC time(s)} & \multicolumn{4}{c|}{SR time(s)} \\ \cline{3-10} & & 10k processes & 40k processes & 90k processes & 160k processes & 10k processes & 40k processes & 90k processes & 160k processes \\ \hline
			\multirow{5}{*}{\begin{tabular}[c]{@{}l@{}}106529\\ parameters\end{tabular}}
			& 8 & 54 & 56 & 56 & 57 & 58 & 41 & 38 & 39 \\
			& 16 & 106 & 106 & 108 & 110 & 66 & 48 & 46 & 47  \\
			& 32 & 202 & 206 & 210 & 208 & 76  & 63  & 59  & 63  \\
			& 64 & 408 & 408 & 410 & 413 & 96 & 95 & 115  &  121  \\
			& 128 & 705 & 702 & 708 & 711 & 135 & 158  & 167  & 213  \\ \hline\hline
			\multirow{4}{*}{\begin{tabular}[c]{@{}l@{}}421953\\ parameters\end{tabular}}
			&8  &  102  &  100  & 104 &  108  & 	666  &  450  &  331  &  300  \\
			&16  &  212  &  212  &  216  &  216  &  746  &  489  &  421  &  433  \\
			&32 &  384  &  386  &  384  &  388  &  909  &  604  &  515  &  540  \\
			&64 &  748  &  750  & 754  & 756 &  1087  &  843  &  766  &  867  \\
			&128 &  1288  &  1280  & 1292  & 1300 &  1459  &  1338  &  1259  &  1521  \\ \hline
		\end{tabular}
	}
	\label{MC-SR-scal}
\end{table*}

The evaluation is conducted on new Sunway supercomputer, and the specific experimental parameters are listed in Table \ref{experimental configuration}. The evaluations from both aspects of computing performance and physics system are elaborated. The many-core CPE optimization for single process and the scalability analysis on the Sunway supercomputer are presented below, as well as the detailed analysis for the parallel CNN-based MCMC and distributed SR computing. Meanwhile, we further evaluate the model training for achieving the global lowest energy, as well as the comparison with state-of-the-art results.

\subsection{Many-core acceleration on SW26010pro}

On the new Sunway supercomputer, most computing performance for each CG is provided by 64 CPEs. In order to achieve high performance on the Sunway supercomputer, it is crucial to effectively use all these CPEs. For the common compute-intensive OPs, swDNN library finally gains 538x, 211x and 505x speedup for Conv2d, Conv2dBackInput and Conv2dBackFilter, respectively. As previously introduced, the neural networks used in this paper involve lots of memory-intensive OPs (i.e., PBC\_padding, Conv1d). Therefore, the special models proposed for physics are much restricted by the gap between the computing capability and the memory access bandwidth. Here we separately optimize these operators on new Sunway heterogeneous processors, especially for several memory-intensive operators.

\begin{table}
\centering
\caption{The many-core acceleration effects for the memory-intensive operations.}
\label{many-core-perf}
\resizebox{8.3cm}{!}{
\begin{tabular}{|c|c|c|c|c|c|}
\hline
\multicolumn{2}{|c|}{Description}                                                                        & SW\_Tile             & SW\_Slice            & SW\_PBC              & SW\_PBC\_grad         \\
\hline
\multirow{5}{*}{\begin{tabular}[c]{@{}c@{}}Testcase\\ input\\ shape\end{tabular}} & \#1                  & {[}32,10,10,128]     & {[}64,100,100,32]    & {[}64,100,32]        & {[}64,144,32]         \\
\cline{2-6}
                                                                                  & \#2                  & {[}32,10,10,256]     & {[}64,100,100,64]    & {[}64,100,64]        & {[}64,144,64]         \\
\cline{2-6}
                                                                                  & \#3                  & {[}32,16,16,128]     & {[}64,256,256,32]    & {[}64,256,32]        & {[}64,400,32]         \\
\cline{2-6}
                                                                                  & \#4                  & {[}32,32,32,64]      & {[}128,100,100,32]   & {[}128,100,32]       & {[}128,144,32]        \\
\cline{2-6}
                                                                                  & \#5                  & {[}64,32,32,256]     & {[}128,256,256,32]   & {[}128,100,64]       & {[}128,144,64]        \\
\hline
\multicolumn{1}{l}{}                                                              & \multicolumn{1}{l}{} & \multicolumn{1}{l}{} & \multicolumn{1}{l}{} & \multicolumn{1}{l}{} & \multicolumn{1}{l}{}  \\
\hline
\multicolumn{2}{|c|}{Testcase}                                                                           & \multicolumn{4}{c|}{Speedup over MPE version}                                              \\
\hline
\multicolumn{2}{|c|}{ \#1}                                                                               & 33.96                & 44.85                & 31.62                & 50.46                 \\
\hline
\multicolumn{2}{|c|}{\#2}                                                                                & 34.45                & 39.5                 & 31.3                 & 53.17                 \\
\hline
\multicolumn{2}{|c|}{\#3}                                                                                & 33.21                & 38.9                 & 33.37                & 60.13                 \\
\hline
\multicolumn{2}{|c|}{\#4}                                                                                & 34.68                & 39.07                & 32.31                & 52.45                 \\
\hline
\multicolumn{2}{|c|}{\#5}                                                                                & 35.93                & 34.38                & 31.41                & 47.65                 \\
\hline
\end{tabular}
}
\end{table}


Table \ref{many-core-perf} presents the test cases and results for the typical memory-intensive operators. Compared to the MPE baseline version, swHMAE achieves great performance improvements. The many-core speedup ratio reaches 34.45 for Tile, 39.34 for Slice, 32 for PBC, and 52.77 for GradPBC. The GradPBC operator obtains better speedup, and it originates from its computing feature. It is not a pure memory access operator, but includes a certain amount of accumulation calculation, which increases the utilization of CPEs. With these highly optimized kernels, the overall speedup for hotspot energy calculation achieves considerable performance improvement over the original MPE version. The many-core accelerated version obtains 90x and 130x speedup for the CNN model with 106529 and 421953 parameters, respectively.

\subsection{Scaling}
For the whole application test, we record the total execution time of a complete training iteration, which includes four kernels. The MCMC stage includes the CNN-based random walking and importance sampling, and the SR stage includes building covariance matrix and matrix inversion.

Fig.\ref{innov-2} presents the CNN-based sampling, which is principally independent among all participated processes. There is little communication overhead for increasing processes and chains. In detail, the CNN computation is related to the lattice size ($\sim L^2$ for the proposed spin pairs and $\sim L^2$ for model forward calculation). On the contrary, the SR computing is determined by the number of network parameters and processes, but is not relevant to the lattice sizes.

	\begin{figure}[tbp]
		\centering
		\begin{subfigure}[b]{.23\textwidth}
			\includegraphics[scale=0.18]{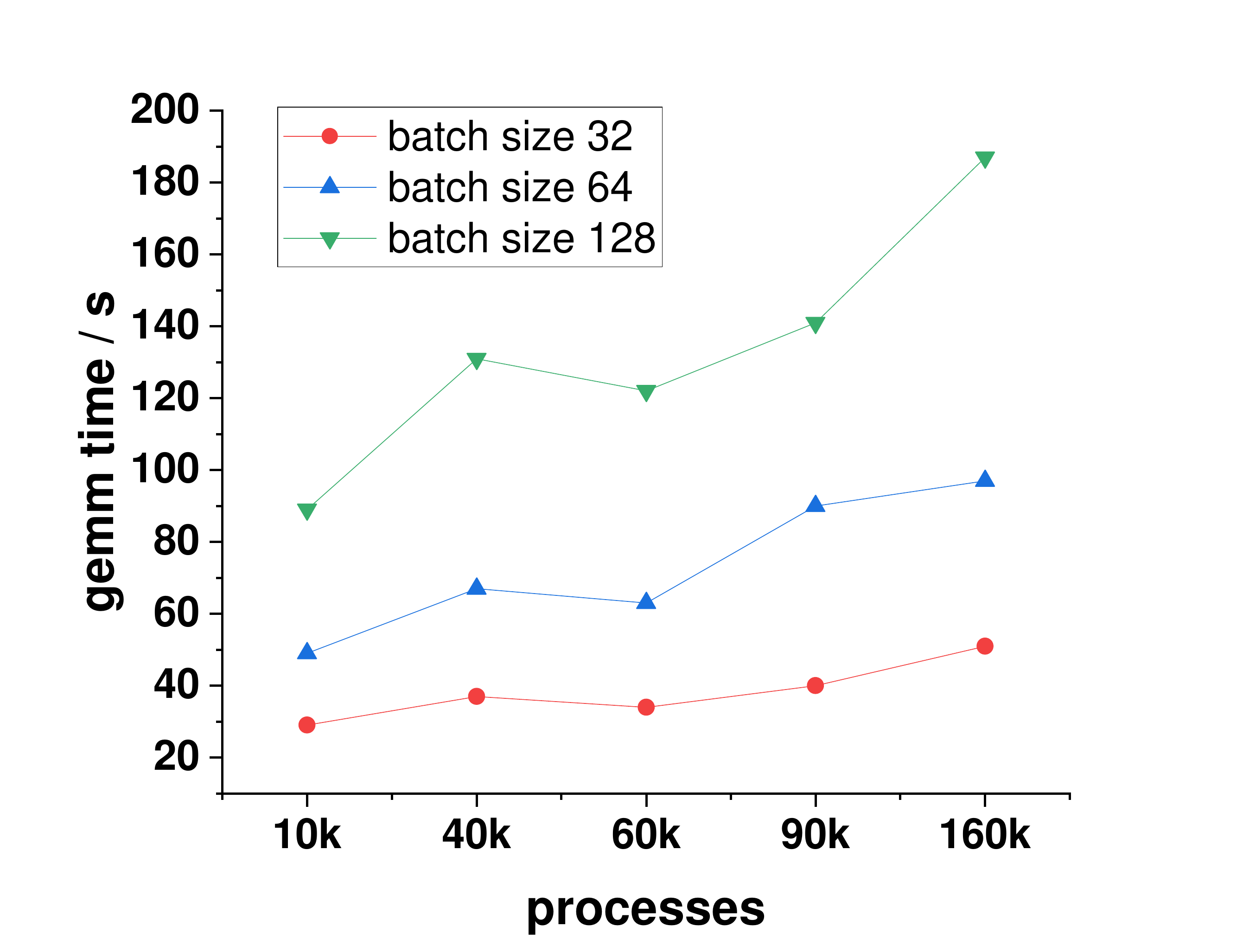}
		\end{subfigure}
		\begin{subfigure}[b]{.23\textwidth}
			\includegraphics[scale=0.18]{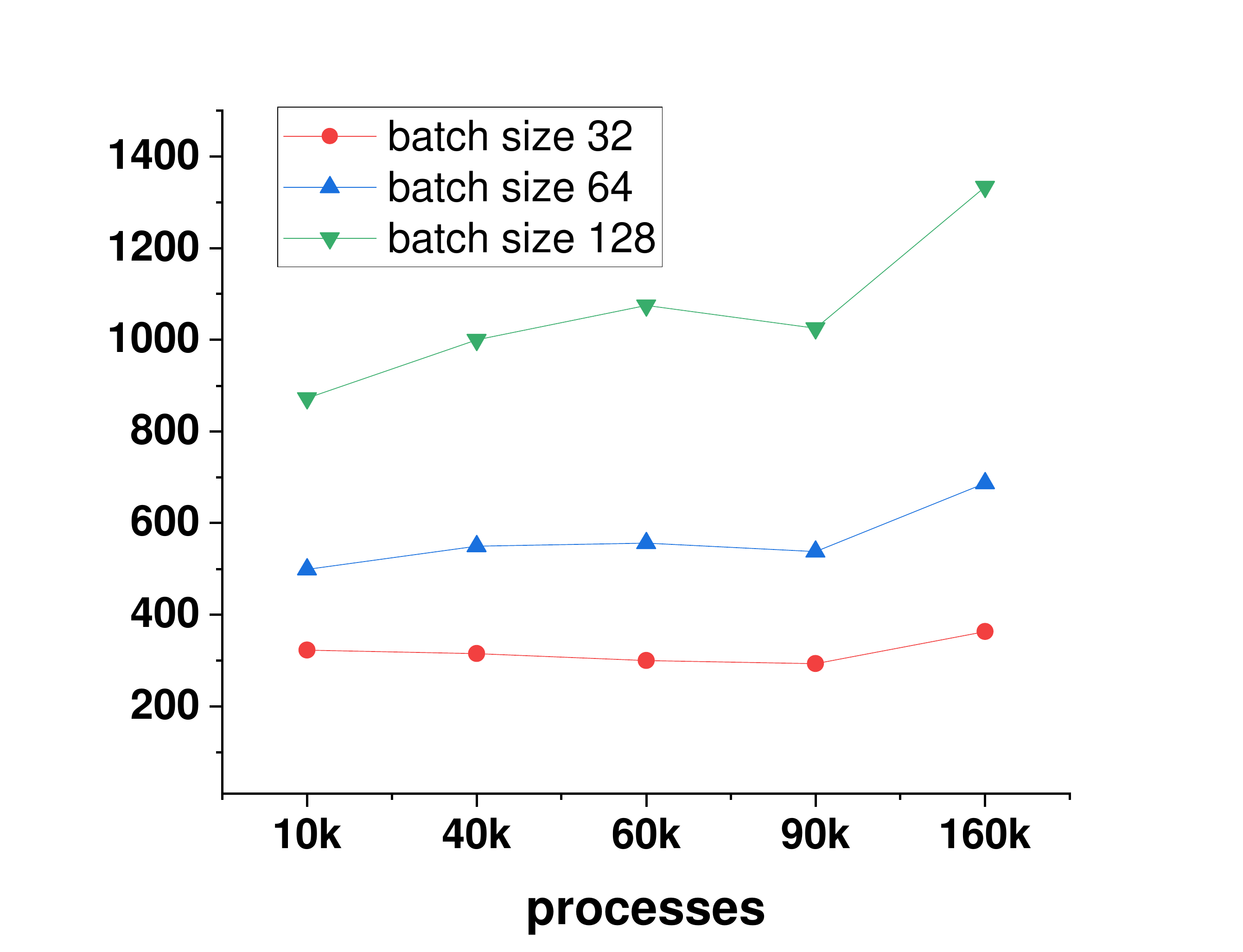}
		\end{subfigure}
    	\begin{subfigure}[b]{.23\textwidth}
    		\includegraphics[scale=0.2]{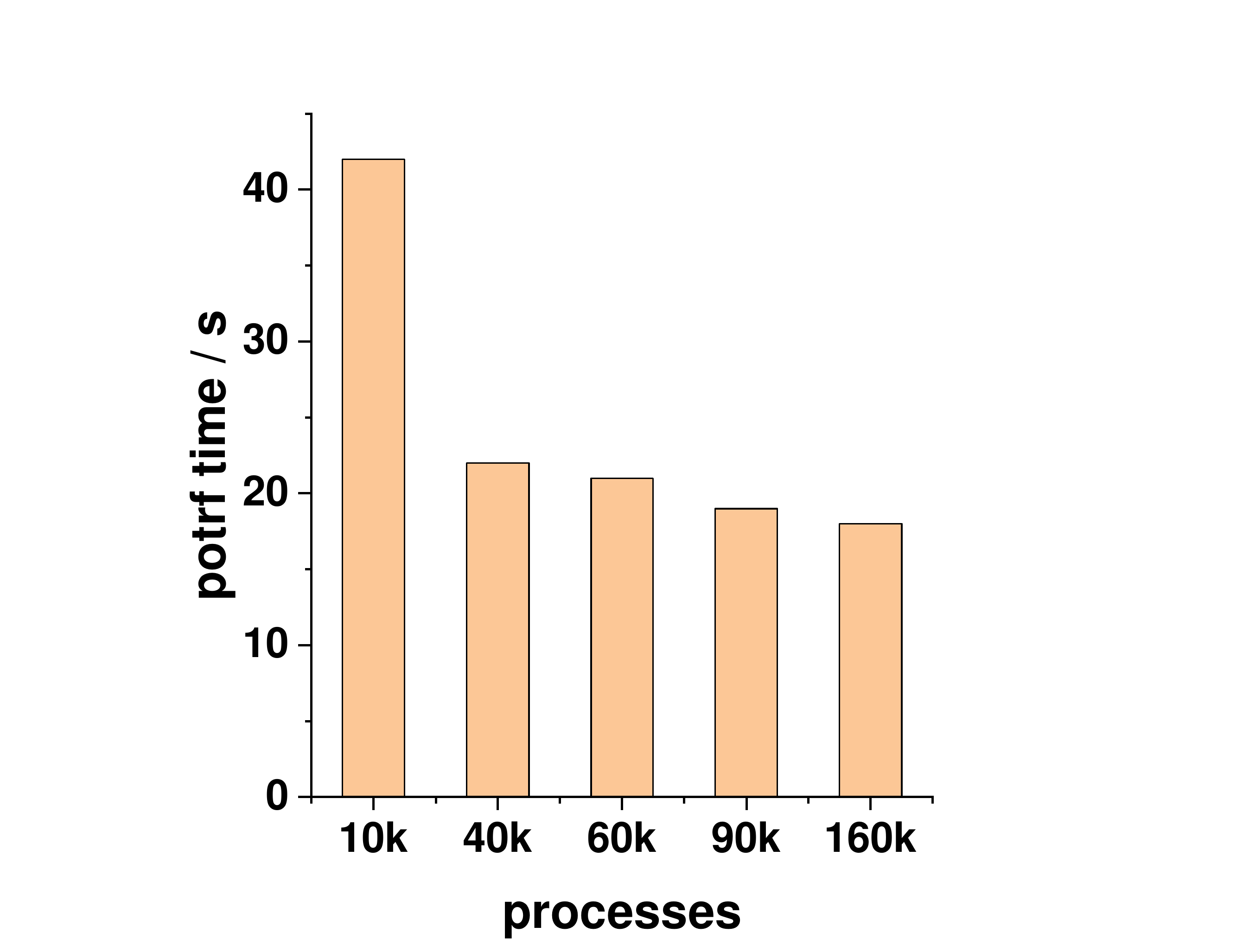}
    	\end{subfigure}
    	\begin{subfigure}[b]{.23\textwidth}
    		\includegraphics[scale=0.2]{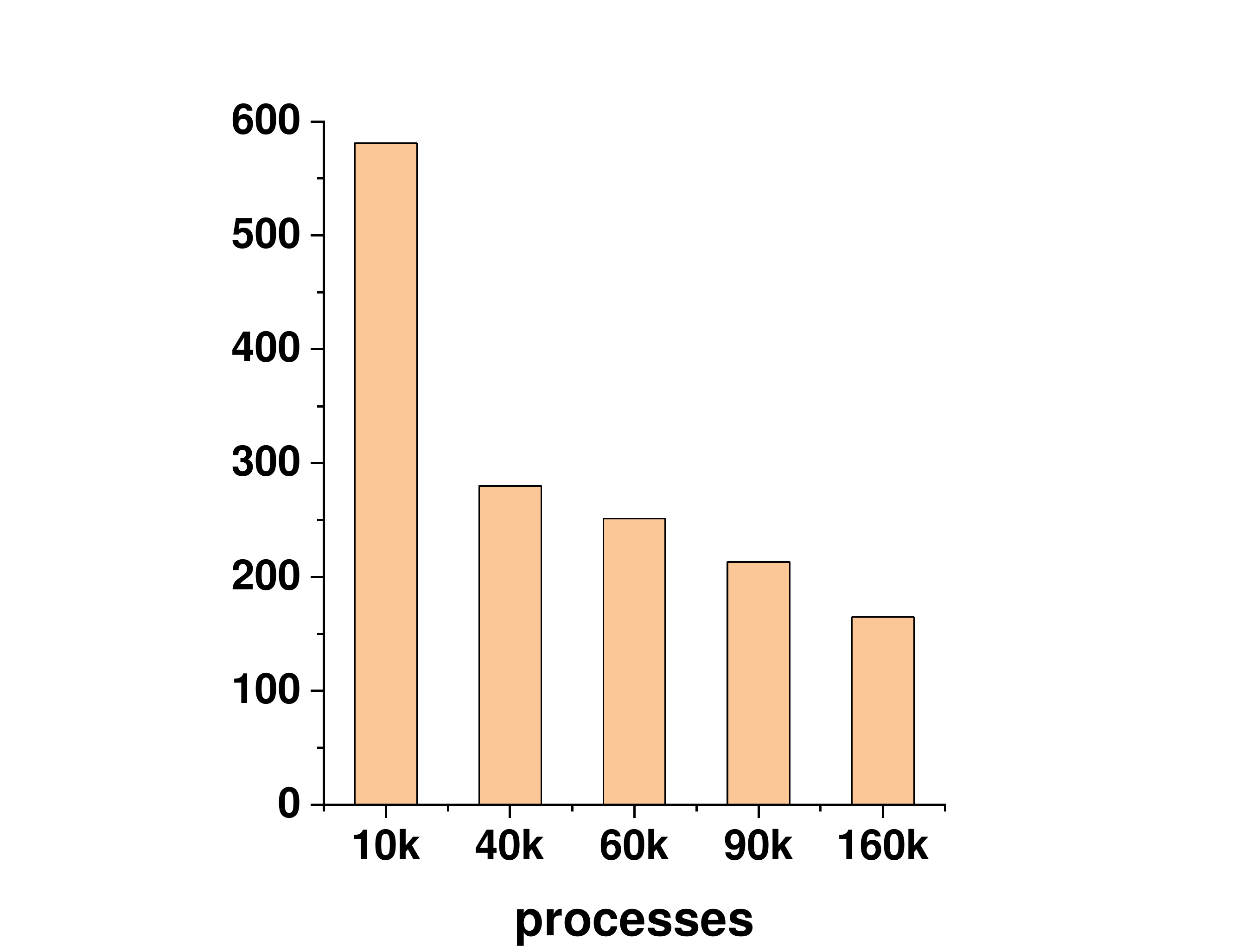}
    	\end{subfigure}
		\caption{Time of gemm (potrf) operation under different number of processes. The matrix size of the left graph is 106529, and that of the right graph is 421953.}
		\label{MCMC-SR}
	\end{figure}
	
\begin{figure*}[htbp]
	\centering
	\begin{subfigure}[b]{.48\textwidth}
		\centerline{\includegraphics[scale=0.29]{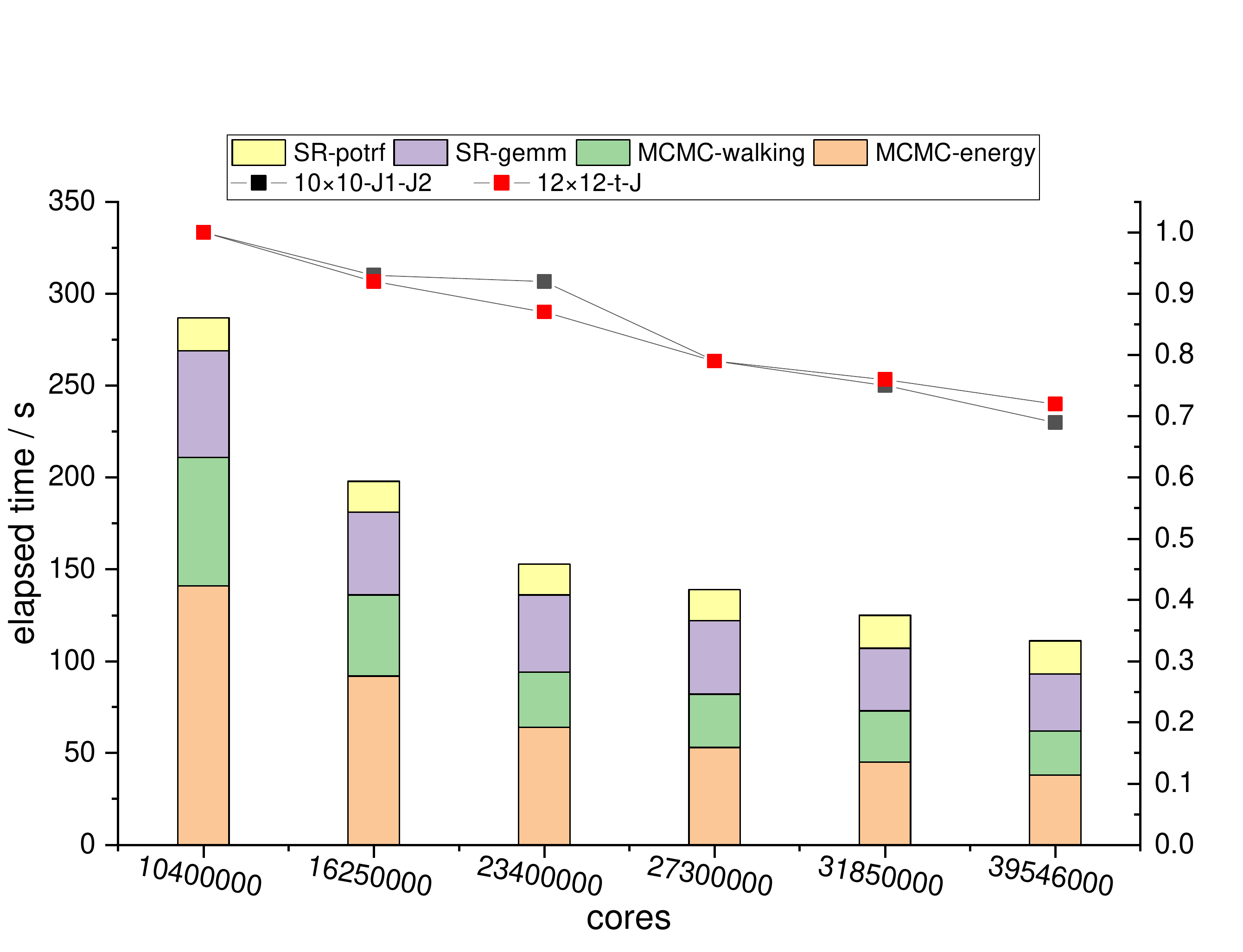}}
		\caption{strong scaling}
	\end{subfigure}
	\begin{subfigure}[b]{.48\textwidth}
		\centerline{\includegraphics[scale=0.29]{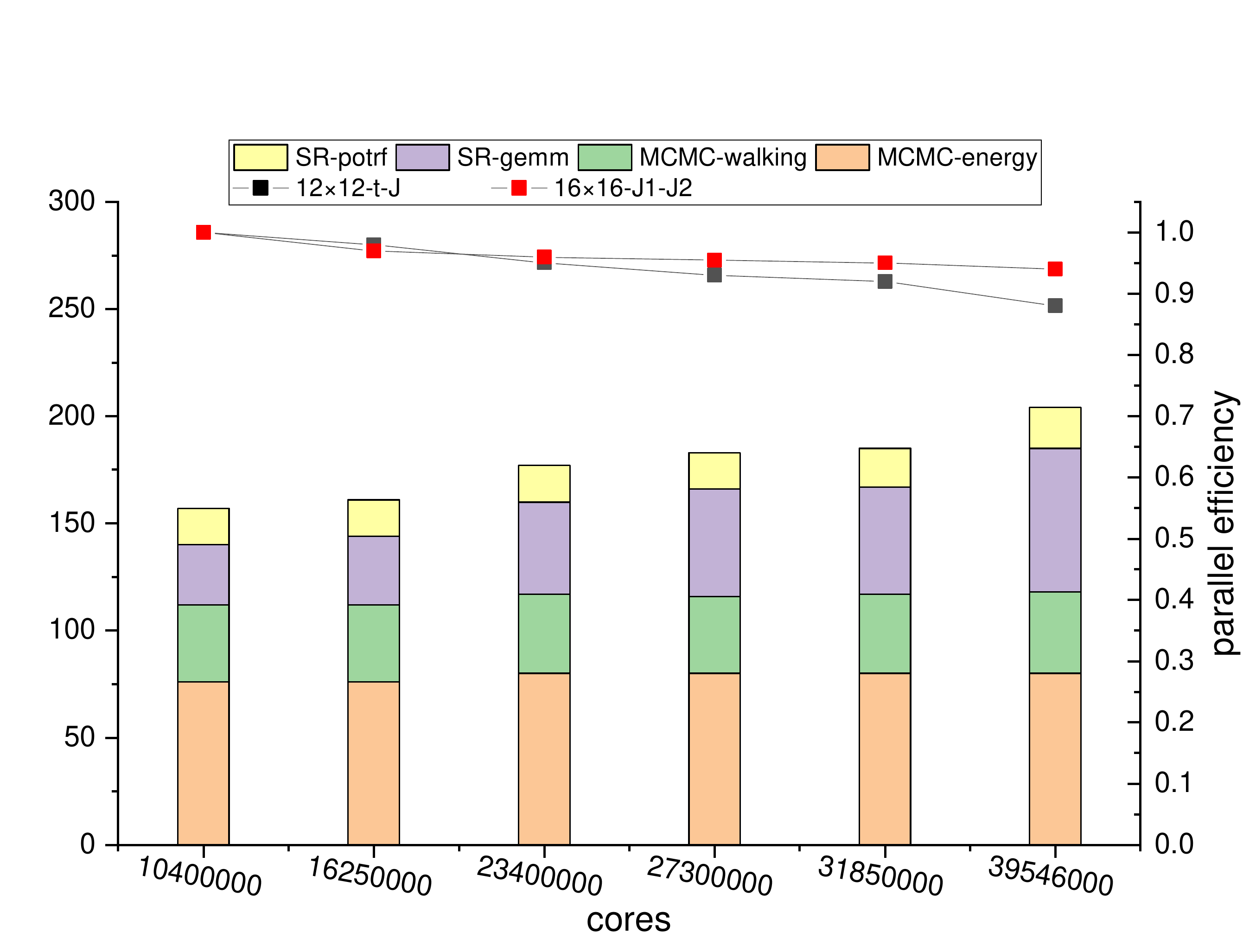}}
		\caption{weak scaling}
	\end{subfigure}
	\caption{Detailed strong scaling (a) and weak scaling (b) results. Two stages (MCMC, SR) with four categories (random walking, importance sampling, matrix multiplication and inversion) are profiled in the time.}
	\label{scaling_detail}
\end{figure*}

\subsubsection{MCMC and SR scalability analysis}
The overall performance results are shown in Table \ref{MC-SR-scal}. The MCMC can be perfectly parallelized and exhibits excellent scalability. There are two hotspots for SR optimization: matrix multiplication (\textit{gemm}) and matrix factorization (\textit{potrf}). The computation of \textit{gemm} increases along with the size of the rows and columns, which are the number of the total Markov chains and parameters. Meanwhile, the computation of \textit{potrf} is only related to the number of parameters, which is completely independent of the system scale or Markov chain number.

For the model with 106529 parameters and small batch size, the computation amount of \textit{potrf} is much higher than that of \textit{gemm}, thus dominating the execution time of SR. The SR time decreases as the processes increase for the parallel \textit{potrf} (Fig. \ref{MCMC-SR}). However, the \textit{gemm} increases (Fig. \ref{MCMC-SR}) as the batch size increases, which gradually dominates the most part of execution time. Since the parallel \textit{potrf} may saturate for the fixed model parameters, the SR time eventually increases with the increasing processes. The 421953-parameter model shows a similar phenomenon, except that the computation amount of \textit{potrf} increases dramatically (\textit{potrf} is sensitive to parameter number), and more processes are required to reach the performance saturation point.

\subsubsection{Strong scaling}

In order to examine the performance of strong scalability, the Markov chains (batch size) per process are decreased in proportion to the process number so that the total Markov chains are kept at around 6 million.



The detailed test results are shown in Fig. \ref{scaling_detail}(a), where the elapsed time of four kernels are introduced. Meanwhile, it demonstrates the parallel efficiency of two test cases with $10 \times 10$ $J1$-$J2$ model and $12\times 12$ $t$-$J$ model, respectively. As is shown in the columns,  the elapsed time of MCMC-energy decreases from 141s to 38s and achieves a parallel efficiency of 97\% when scaling to 39,546,000 cores. The MCMC-walking time decreases from 70s to 24s and achieves 77\% parallel efficiency. The SR part involves distributed and parallel ScaLapack calculation. The performance strongly relies on the communication and the elapsed time decreases from 76s to 49s with a 40\% parallel efficiency. Overall, the $10 \times 10$ case obtains a 2.6x throughput improvement and a parallel efficiency of 69\% when scaling to 39,546,000 cores, while the $12\times 12$ case shows slightly better strong scalability than the $10 \times 10$ case (benefiting from the fixed SR time and increasing proportion of MCMC time), which obtains a 2.8x throughput improvement and 72\% parallel efficiency.

\subsubsection{Weak scaling}

The detailed weak scalability results are drawn in Fig. \ref{scaling_detail}(b), including the elapsed time of MCMC (random walking, energy computation) and SR (covariance matrix, and matrix inversion), as well as the parallel efficiency of 2 test cases with $12 \times 12$ $t$-$J$ model and $16\times 16$ $J1$-$J2$ model, respectively, when scaling to 39,546,000 cores. As is shown in the columns, when scaling from 10,400,000 cores to 39,546,000 cores, the elapsed time of MCMC-energy and MCMC-waling is almost unchanged, showing excellent weak scalability (almost linear scalable). As for SR computing, the communication pressure for ScaLapack grows as the process increases, and this communication overhead affects the scalability of the SR part to some extent. Nevertheless, MCMC process takes a major part of the total execution time, making our application as a whole with satisfactory weak scalability. For the case with $L$=12 lattice size, it scales up to nearly 40 million cores with a parallel efficiency of 88\% and a throughput of nearly 50,000 Markov chains per second, as well as over 10 million Markov chains in total; and for the $L$=16 case, it achieves better result with 94\% parallel efficiency.

\subsection{Simulation and Validation}

\subsubsection{Transfer learning of $J1$-$J2$ model}
The results of transferring learning are denoted by the CNN1 with 106529 parameters. When $L$=6, the energy is decreased to -0.5002 after 100 SR steps. Then the CNN1 is transferred to $L$=10 lattice, and energy is decreased from -0.4673 to -0.4961 after 350 SR steps. The high scalability of CNN1 significantly reduces the optimization difficulty for large lattices. The lattice is finally increased to $L$=36 and the corresponding initial energy is -0.49585.


%

\subsubsection{Initial state selection of $t$-$J$ model}
\begin{table}[htbp]
	\centering
	\caption{Average energies of the initial states found for the $t$-$J$ model with different numbers of Markov chains.}
	\resizebox{8.3cm}{!}{
		\begin{tabular}{|l|c|c|c|c|}
			\hline
			\diagbox{chains}{size} & 8×8 & 12×12 & 16×16 & 24×24 \\
			\hline
			100 & -0.2171 & -0.1787 & -0.1803 & -0.1560 \\
			\hline
			1000 & -0.2929 & -0.2845 & -0.2792 & -0.2640 \\
			\hline
			10000 & -0.3388 & -0.3348 & -0.3152 & -0.3129 \\
			\hline
			100000 & -0.3590 & -0.3446 & -0.3365 & -0.3298 \\
			\hline
		\end{tabular}
	}
	\label{initial_state}
\end{table}

Table \ref{initial_state} presents the average energies of the best initial states found via different number of Markov chains. The energies are achieved by the CNN2 with 113815 parameters. Obviously, the larger the number of Markov chains involved in the parallel search, the higher the quality of the initial states found. Because of the large Hilbert space, the difficulty of finding high-quality initial states for large lattice is significantly higher than that for small lattice. However the difficulty can be overcome by increasing chain number. For small lattice quantum systems, the quality of the initial state has relatively little influence on energy convergence. 
With the selected initial states, after 100 SR steps, the energy decreases from -0.3517 to -0.4523 for $12\times 12$ lattice; from -0.3483 to -0.4599 for $16\times 16$ lattice, and from -0.2900 to -0.4173 for $24\times 24$ lattice.

\subsection{Optimization Results}
\begin{figure}[htbp]
	    \centerline{\includegraphics[scale=0.5]{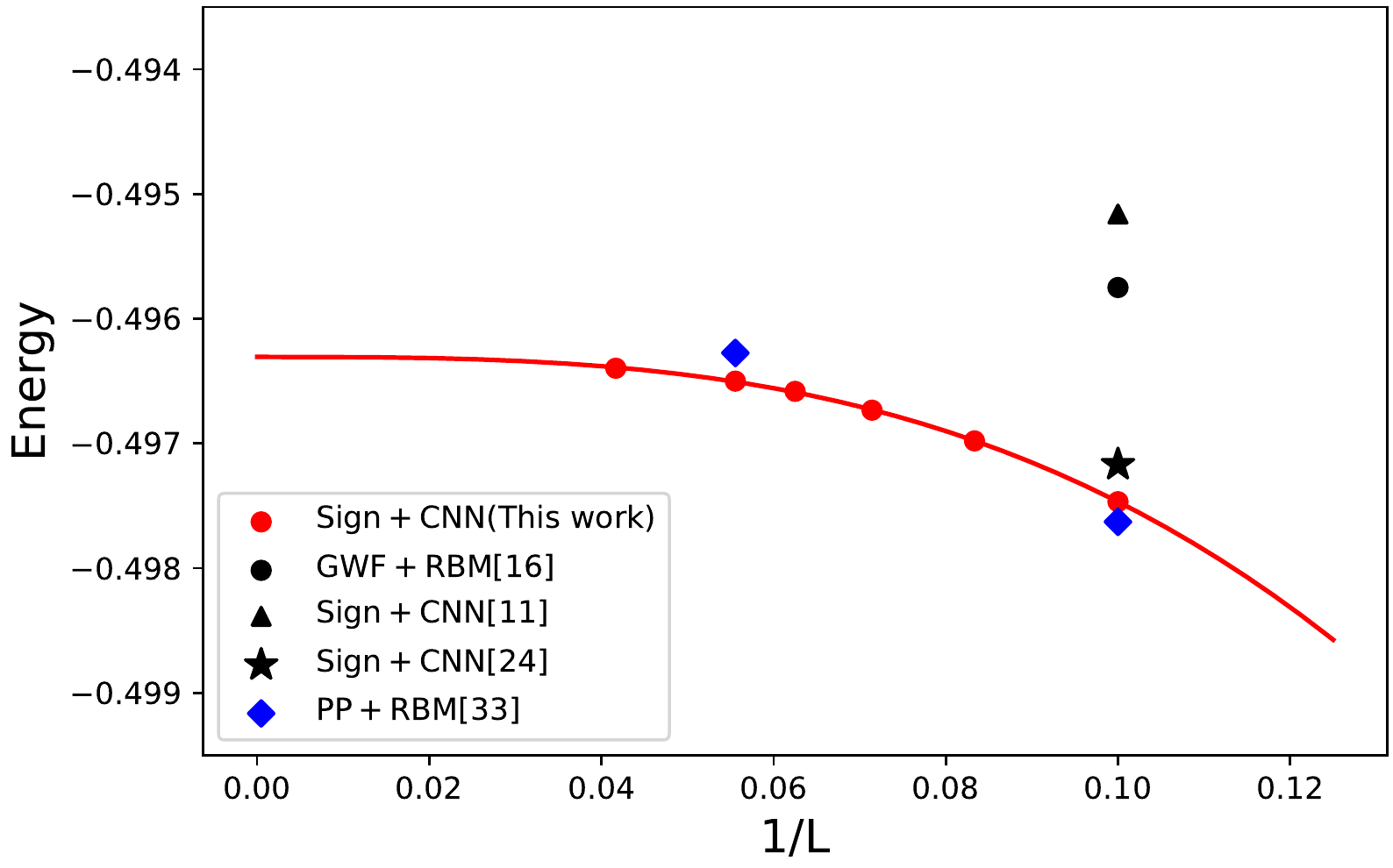}}
	    \caption{Comparing the ground energies of the $J1$-$J2$ model on different lattice size $L$,
       obtained by CNN1 to the current state-of-art results. }
	    \label{J20.5_E}
\end{figure}
\begin{table}[]
\caption{Comparison of  the ground state energies of the $t$-$J$ model, obtained by CNN2 with PBC, and PEPS++ with OBC }
\label{tJ_E}
\begin{tabular}{ccc}
\hline
\multicolumn{1}{|c|}{L}  & \multicolumn{1}{c|}{CNN2(PBC)}      & \multicolumn{1}{c|}{PEPS(OBC)}      \\ \hline
\multicolumn{1}{|c|}{8}  & \multicolumn{1}{c|}{-0.60965} & \multicolumn{1}{c|}{-0.61184} \\
\multicolumn{1}{|c|}{12} & \multicolumn{1}{c|}{-0.63217} & \multicolumn{1}{c|}{-0.62973} \\ \hline
                         &                               &
\end{tabular}
\end{table}
The optimized  energies of the $J1$-$J2$ model and the $t$-$J$ model are presented.
For the $J1$-$J2$ model in the maximal frustration region $J_2=0.5J_1$, the convergence of energy with respect to lattice size $L$ is depicted in Fig.\ref{J20.5_E}. In the figure, the energies are obtained by the CNN1 with the parameter number of 106529. The solid line is fitted with the function $E=a/L^3+b$. Other state-of-art results are also shown in the figure for comparison. Comparing the energies in this work to the results obtained by the PP+RBM method on the ``Fugaku" supercomputer\cite{nomura2021dirac}. When $L$=10 the energy in this work is -0.497468, which is $3.2\times 10^{-4}$ higher than the -0.497629 reported in \cite{nomura2021dirac}, and $5.8\times 10^{-4}$ lower than the -0.49717 reported in \cite{li2022bridging}. For $L$=18, the ground state energy obtained in this work is -0.496500, which is $4.5\times 10^{-4}$ lower than -0.496275 reported in \cite{nomura2021dirac}. Thanks to excellent scalability of the CNN1, we are able to study the model on the sizes that are significantly larger than previous works.

The ground state energies of the  $t$-$J$ model are presented in Table.\ref{tJ_E}. Unfortunately, so far there is no reference ground state energies for the  $t$-$J$ with PBC, we therefore compare the results to those of OBC calculated via PEPS in Ref.\cite{PhysRevB.99.195153}.
In the table, the energies are obtained by the CNN2 with the parameter number of 113815. When $L$=8, the energy achieved by CNN2 is -0.60965, which is $3\times 10^{-3}$ higher than that reported in \cite{PhysRevB.99.195153}. When $L$=12, the energy achieved by CNN2 is -0.63217, which is $3.8\times 10^{-3}$ lower than that reported in \cite{PhysRevB.99.195153}. We would like to note that the energies of different boundary conditions can not be compared directly, which may have some small difference. The results show here are only a
demonstration of  the effectiveness of the CNN2 for Fermion systems.

\section{IMPLICATIONS}

In this work, we report simulations of challenging quantum many-body problems
 that combines the state-of-art deep-learning method and efficient implementation on the next-generation computational platform.
 We have obtained the ground state energies of the $J_1$-$J_2$ model and $t$-$J$ model with accuracy
 that surpass the of current-state-of-art results.
 With this ability, we are able to simulate a large class of important physical models to the high accuracy
to  make confident conclusions about the physics.
The method therefore opens up a new promising path to solve the
extremely important and challenging problems that have plagued people for more than half a century,
and would help us to understand exotic physical phenomena, such as quantum spin liquids ,
high temperature superconductivity~\cite{Dagotto1994},
supersolid~\cite{Boninsegni2012}, heavy fermions~\cite{Steppke2013}, fractional quantum Hall effects, \cite{Tsui1982,Laugh1983,Stormer1999}
and many more, which is not only essential for the understanding
of fundamental physics but also have many important applications.

On the other hand, this work demonstrated the effectiveness of solving the quantum many-body problem via
the deep learning methods on the supercomputer platform, which is completely different from traditional machine learning tasks.
These applications represent a large class of scientific problems that
are more challenging in the representation ability of the network and the accuracy requirements of the results.
It also puts forward higher requirements for efficient optimization algorithm and implementation.
This framework may apply to other problems, e.g., the simulation of quantum circuit on classic supercomputers.
Currently tensor network methods have shown high efficiency for quantum circuit simulations on supercomputers\cite{GB2021}
on the lattice as large as $10 \times 10$. The scalable neural network based quantum state representation
may capable to perform quantum circuit simulations for even larger number of qubits.

Our framework is not limited to Sunway, and it can be easily ported to other supercomputing platforms. However, as computational power increases, the overall application performance will be limited by the gap between the computing capability and the memory access bandwidth. Here we take the computational power v.s. memory bandwidth ratio (FLOP/Byte) to measure the potential performance. Because the single precision does not meet the requirements of achieving faithful ground state energies, here we focus on the double precision FLOP/Byte. With the double precision support of tensor cores, the FLOP/Byte ratio on A100 GPU is 9.56 (19.5TFLOPS/2039GB/s=9.56FLOP/Byte). The double precision FLOP/Byte ratio on Fujitsu A64FX CPU is 3.3 (3.3792TFLOPS/1024GB/s=3.3FLOP/Byte). Considering the performance of the LDM used in sw26010pro is not competitive to the performance of HBM2 used in A100 and A64FX, therefore the performance of our framework can be further boosted on Summit or Fugaku systems.

\section*{Acknowledgment}


\bibliographystyle{ACM-Reference-Format}
\bibliography{references,references2}


\begin{thebibliography}{50}


\ifx \showCODEN    \undefined \def \showCODEN     #1{\unskip}     \fi
\ifx \showDOI      \undefined \def \showDOI       #1{#1}\fi
\ifx \showISBNx    \undefined \def \showISBNx     #1{\unskip}     \fi
\ifx \showISBNxiii \undefined \def \showISBNxiii  #1{\unskip}     \fi
\ifx \showISSN     \undefined \def \showISSN      #1{\unskip}     \fi
\ifx \showLCCN     \undefined \def \showLCCN      #1{\unskip}     \fi
\ifx \shownote     \undefined \def \shownote      #1{#1}          \fi
\ifx \showarticletitle \undefined \def \showarticletitle #1{#1}   \fi
\ifx \showURL      \undefined \def \showURL       {\relax}        \fi
\providecommand\bibfield[2]{#2}
\providecommand\bibinfo[2]{#2}
\providecommand\natexlab[1]{#1}
\providecommand\showeprint[2][]{arXiv:#2}

\bibitem[\protect\citeauthoryear{??}{qua}{2018}]%
        {quantumhalleffect_url}
 \bibinfo{year}{2018}\natexlab{}.
\newblock \bibinfo{booktitle}{\emph{The Fractional Quantum Hall Effect}}.
\newblock
\urldef\tempurl%
\url{https://www.nobelprize.org/uploads/2018/06/stormer-lecture.pdf}
\showURL{%
\tempurl}


\bibitem[\protect\citeauthoryear{??}{sup}{2018}]%
        {superconductivity_url}
 \bibinfo{year}{2018}\natexlab{}.
\newblock \bibinfo{booktitle}{\emph{Nobel Laureates in Superconductivity}}.
\newblock
\urldef\tempurl%
\url{http://past.ieeecsc.org/pages/nobel-laureates-superconductivity}
\showURL{%
\tempurl}


\bibitem[\protect\citeauthoryear{Anderson}{Anderson}{2007}]%
        {Anderson2007}
\bibfield{author}{\bibinfo{person}{J.~B Anderson}.}
  \bibinfo{year}{2007}\natexlab{}.
\newblock \bibinfo{booktitle}{\emph{Quantum Monte Carlo: Origins, Development,
  Applications}}.
\newblock \bibinfo{publisher}{Oxford University Press}.
\newblock


\bibitem[\protect\citeauthoryear{Batista and Ortiz}{Batista and Ortiz}{2001}]%
        {JW}
\bibfield{author}{\bibinfo{person}{C.~D. Batista} {and} \bibinfo{person}{G.
  Ortiz}.} \bibinfo{year}{2001}\natexlab{}.
\newblock \showarticletitle{Generalized Jordan-Wigner Transformations}.
\newblock \bibinfo{journal}{\emph{Phys. Rev. Lett.}}  \bibinfo{volume}{86}
  (\bibinfo{date}{Feb} \bibinfo{year}{2001}), \bibinfo{pages}{1082--1085}.
\newblock
Issue 6.
\urldef\tempurl%
\url{https://doi.org/10.1103/PhysRevLett.86.1082}
\showDOI{\tempurl}


\bibitem[\protect\citeauthoryear{Boninsegni and Prokof'ev}{Boninsegni and
  Prokof'ev}{2012}]%
        {Boninsegni2012}
\bibfield{author}{\bibinfo{person}{M Boninsegni} {and} \bibinfo{person}{N.~V.
  Prokof'ev}.} \bibinfo{year}{2012}\natexlab{}.
\newblock \showarticletitle{Supersolids: What and where are they?}
\newblock \bibinfo{journal}{\emph{Rev. Mod. Phys.}}  \bibinfo{volume}{84}
  (\bibinfo{year}{2012}), \bibinfo{pages}{759--776}.
\newblock
Issue 2.


\bibitem[\protect\citeauthoryear{Broholm, Cava, Kivelson, Nocera, Norman, and
  Senthil}{Broholm et~al\mbox{.}}{2020}]%
        {QSL}
\bibfield{author}{\bibinfo{person}{C. Broholm}, \bibinfo{person}{R.~J. Cava},
  \bibinfo{person}{S.~A. Kivelson}, \bibinfo{person}{D.~G. Nocera},
  \bibinfo{person}{M.~R. Norman}, {and} \bibinfo{person}{T. Senthil}.}
  \bibinfo{year}{2020}\natexlab{}.
\newblock \showarticletitle{Quantum spin liquids}.
\newblock \bibinfo{journal}{\emph{Science}} \bibinfo{volume}{367},
  \bibinfo{number}{6475} (\bibinfo{year}{2020}), \bibinfo{pages}{eaay0668}.
\newblock
\urldef\tempurl%
\url{https://doi.org/10.1126/science.aay0668}
\showDOI{\tempurl}
\showeprint{https://www.science.org/doi/pdf/10.1126/science.aay0668}


\bibitem[\protect\citeauthoryear{Carleo, Cirac, Cranmer, Daudet, Schuld,
  Tishby, Vogt-Maranto, and Zdeborov\'a}{Carleo et~al\mbox{.}}{2019}]%
        {ML_physics_review1}
\bibfield{author}{\bibinfo{person}{Giuseppe Carleo}, \bibinfo{person}{Ignacio
  Cirac}, \bibinfo{person}{Kyle Cranmer}, \bibinfo{person}{Laurent Daudet},
  \bibinfo{person}{Maria Schuld}, \bibinfo{person}{Naftali Tishby},
  \bibinfo{person}{Leslie Vogt-Maranto}, {and} \bibinfo{person}{Lenka
  Zdeborov\'a}.} \bibinfo{year}{2019}\natexlab{}.
\newblock \showarticletitle{Machine learning and the physical sciences}.
\newblock \bibinfo{journal}{\emph{Rev. Mod. Phys.}}  \bibinfo{volume}{91}
  (\bibinfo{date}{Dec} \bibinfo{year}{2019}), \bibinfo{pages}{045002}.
\newblock
Issue 4.
\urldef\tempurl%
\url{https://doi.org/10.1103/RevModPhys.91.045002}
\showDOI{\tempurl}


\bibitem[\protect\citeauthoryear{Carleo and Troyer}{Carleo and Troyer}{2017}]%
        {carleo2017solving}
\bibfield{author}{\bibinfo{person}{Giuseppe Carleo} {and}
  \bibinfo{person}{Matthias Troyer}.} \bibinfo{year}{2017}\natexlab{}.
\newblock \showarticletitle{Solving the quantum many-body problem with
  artificial neural networks}.
\newblock \bibinfo{journal}{\emph{Science}} \bibinfo{volume}{355},
  \bibinfo{number}{6325} (\bibinfo{year}{2017}), \bibinfo{pages}{602--606}.
\newblock


\bibitem[\protect\citeauthoryear{Carrasquilla}{Carrasquilla}{2020}]%
        {ML_physics_review2}
\bibfield{author}{\bibinfo{person}{Juan Carrasquilla}.}
  \bibinfo{year}{2020}\natexlab{}.
\newblock \showarticletitle{Machine learning for quantum matter}.
\newblock \bibinfo{journal}{\emph{Advances in Physics: X}} \bibinfo{volume}{5},
  \bibinfo{number}{1} (\bibinfo{year}{2020}), \bibinfo{pages}{1797528}.
\newblock
\urldef\tempurl%
\url{https://doi.org/10.1080/23746149.2020.1797528}
\showDOI{\tempurl}
\showeprint{https://doi.org/10.1080/23746149.2020.1797528}


\bibitem[\protect\citeauthoryear{Choi, Dongarra, Pozo, and Walker}{Choi
  et~al\mbox{.}}{1992}]%
        {choi1992scalapack}
\bibfield{author}{\bibinfo{person}{Jaeyoung Choi}, \bibinfo{person}{Jack~J
  Dongarra}, \bibinfo{person}{Roldan Pozo}, {and} \bibinfo{person}{David~W
  Walker}.} \bibinfo{year}{1992}\natexlab{}.
\newblock \showarticletitle{ScaLAPACK: A scalable linear algebra library for
  distributed memory concurrent computers}. In \bibinfo{booktitle}{\emph{The
  Fourth Symposium on the Frontiers of Massively Parallel Computation}}. IEEE
  Computer Society, \bibinfo{pages}{120--121}.
\newblock


\bibitem[\protect\citeauthoryear{Choo, Neupert, and Carleo}{Choo
  et~al\mbox{.}}{2019}]%
        {choo2019two}
\bibfield{author}{\bibinfo{person}{Kenny Choo}, \bibinfo{person}{Titus
  Neupert}, {and} \bibinfo{person}{Giuseppe Carleo}.}
  \bibinfo{year}{2019}\natexlab{}.
\newblock \showarticletitle{Two-dimensional frustrated J 1- J 2 model studied
  with neural network quantum states}.
\newblock \bibinfo{journal}{\emph{Physical Review B}} \bibinfo{volume}{100},
  \bibinfo{number}{12} (\bibinfo{year}{2019}), \bibinfo{pages}{125124}.
\newblock


\bibitem[\protect\citeauthoryear{Dagotto}{Dagotto}{1994}]%
        {Dagotto1994}
\bibfield{author}{\bibinfo{person}{E. Dagotto}.}
  \bibinfo{year}{1994}\natexlab{}.
\newblock \showarticletitle{Correlated electrons in high-temperature
  superconductors}.
\newblock \bibinfo{journal}{\emph{Rev. Mod. Phys.}}  \bibinfo{volume}{66}
  (\bibinfo{year}{1994}), \bibinfo{pages}{763--840}.
\newblock
Issue 3.


\bibitem[\protect\citeauthoryear{Degrave, Felici, Buchli, Neunert, Tracey,
  Carpanese, Ewalds, Hafner, Abdolmaleki, de~Las~Casas, et~al\mbox{.}}{Degrave
  et~al\mbox{.}}{2022}]%
        {degrave2022magnetic}
\bibfield{author}{\bibinfo{person}{Jonas Degrave}, \bibinfo{person}{Federico
  Felici}, \bibinfo{person}{Jonas Buchli}, \bibinfo{person}{Michael Neunert},
  \bibinfo{person}{Brendan Tracey}, \bibinfo{person}{Francesco Carpanese},
  \bibinfo{person}{Timo Ewalds}, \bibinfo{person}{Roland Hafner},
  \bibinfo{person}{Abbas Abdolmaleki}, \bibinfo{person}{Diego de Las~Casas},
  {et~al\mbox{.}}} \bibinfo{year}{2022}\natexlab{}.
\newblock \showarticletitle{Magnetic control of tokamak plasmas through deep
  reinforcement learning}.
\newblock \bibinfo{journal}{\emph{Nature}} \bibinfo{volume}{602},
  \bibinfo{number}{7897} (\bibinfo{year}{2022}), \bibinfo{pages}{414--419}.
\newblock


\bibitem[\protect\citeauthoryear{Dong, Wang, Han, Guo, and He}{Dong
  et~al\mbox{.}}{2019}]%
        {PhysRevB.99.195153}
\bibfield{author}{\bibinfo{person}{Shao-Jun Dong}, \bibinfo{person}{Chao Wang},
  \bibinfo{person}{Yongjian Han}, \bibinfo{person}{Guang-can Guo}, {and}
  \bibinfo{person}{Lixin He}.} \bibinfo{year}{2019}\natexlab{}.
\newblock \showarticletitle{Gradient optimization of fermionic projected
  entangled pair states on directed lattices}.
\newblock \bibinfo{journal}{\emph{Phys. Rev. B}}  \bibinfo{volume}{99}
  (\bibinfo{date}{May} \bibinfo{year}{2019}), \bibinfo{pages}{195153}.
\newblock
Issue 19.
\urldef\tempurl%
\url{https://doi.org/10.1103/PhysRevB.99.195153}
\showDOI{\tempurl}


\bibitem[\protect\citeauthoryear{Fang, Fu, Zhao, Chen, Zheng, and Yang}{Fang
  et~al\mbox{.}}{2017}]%
        {fang2017swdnn}
\bibfield{author}{\bibinfo{person}{Jiarui Fang}, \bibinfo{person}{Haohuan Fu},
  \bibinfo{person}{Wenlai Zhao}, \bibinfo{person}{Bingwei Chen},
  \bibinfo{person}{Weijie Zheng}, {and} \bibinfo{person}{Guangwen Yang}.}
  \bibinfo{year}{2017}\natexlab{}.
\newblock \showarticletitle{swdnn: A library for accelerating deep learning
  applications on sunway taihulight}. In \bibinfo{booktitle}{\emph{2017 IEEE
  International Parallel and Distributed Processing Symposium (IPDPS)}}. IEEE,
  \bibinfo{pages}{615--624}.
\newblock


\bibitem[\protect\citeauthoryear{Ferrari, Becca, and Carrasquilla}{Ferrari
  et~al\mbox{.}}{2019}]%
        {ferrari2019neural}
\bibfield{author}{\bibinfo{person}{Francesco Ferrari},
  \bibinfo{person}{Federico Becca}, {and} \bibinfo{person}{Juan Carrasquilla}.}
  \bibinfo{year}{2019}\natexlab{}.
\newblock \showarticletitle{Neural Gutzwiller-projected variational wave
  functions}.
\newblock \bibinfo{journal}{\emph{Physical Review B}} \bibinfo{volume}{100},
  \bibinfo{number}{12} (\bibinfo{year}{2019}), \bibinfo{pages}{125131}.
\newblock


\bibitem[\protect\citeauthoryear{Han, Jentzen, and E}{Han
  et~al\mbox{.}}{2018}]%
        {doi:10.1073/pnas.1718942115}
\bibfield{author}{\bibinfo{person}{Jiequn Han}, \bibinfo{person}{Arnulf
  Jentzen}, {and} \bibinfo{person}{Weinan E}.} \bibinfo{year}{2018}\natexlab{}.
\newblock \showarticletitle{Solving high-dimensional partial differential
  equations using deep learning}.
\newblock \bibinfo{journal}{\emph{Proceedings of the National Academy of
  Sciences}} \bibinfo{volume}{115}, \bibinfo{number}{34}
  (\bibinfo{year}{2018}), \bibinfo{pages}{8505--8510}.
\newblock
\urldef\tempurl%
\url{https://doi.org/10.1073/pnas.1718942115}
\showDOI{\tempurl}
\showeprint{https://www.pnas.org/doi/pdf/10.1073/pnas.1718942115}


\bibitem[\protect\citeauthoryear{He, An, Yang, Wang, Chen, Wang, Liang, Dong,
  Sun, Han, et~al\mbox{.}}{He et~al\mbox{.}}{2018}]%
        {he2018peps++}
\bibfield{author}{\bibinfo{person}{Lixin He}, \bibinfo{person}{Hong An},
  \bibinfo{person}{Chao Yang}, \bibinfo{person}{Fei Wang},
  \bibinfo{person}{Junshi Chen}, \bibinfo{person}{Chao Wang},
  \bibinfo{person}{Weihao Liang}, \bibinfo{person}{Shaojun Dong},
  \bibinfo{person}{Qiao Sun}, \bibinfo{person}{Wenting Han}, {et~al\mbox{.}}}
  \bibinfo{year}{2018}\natexlab{}.
\newblock \showarticletitle{Peps++: towards extreme-scale simulations of
  strongly correlated quantum many-particle models on sunway taihulight}.
\newblock \bibinfo{journal}{\emph{IEEE Transactions on Parallel and Distributed
  Systems}} \bibinfo{volume}{29}, \bibinfo{number}{12} (\bibinfo{year}{2018}),
  \bibinfo{pages}{2838--2848}.
\newblock


\bibitem[\protect\citeauthoryear{Hu, Becca, Parola, and Sorella}{Hu
  et~al\mbox{.}}{2013}]%
        {hu2013direct}
\bibfield{author}{\bibinfo{person}{Wen-Jun Hu}, \bibinfo{person}{Federico
  Becca}, \bibinfo{person}{Alberto Parola}, {and} \bibinfo{person}{Sandro
  Sorella}.} \bibinfo{year}{2013}\natexlab{}.
\newblock \showarticletitle{Direct evidence for a gapless Z 2 spin liquid by
  frustrating N{\'e}el antiferromagnetism}.
\newblock \bibinfo{journal}{\emph{Physical Review B}} \bibinfo{volume}{88},
  \bibinfo{number}{6} (\bibinfo{year}{2013}), \bibinfo{pages}{060402}.
\newblock


\bibitem[\protect\citeauthoryear{Inui, Kato, and Motome}{Inui
  et~al\mbox{.}}{2021}]%
        {inui2021determinant}
\bibfield{author}{\bibinfo{person}{Koji Inui}, \bibinfo{person}{Yasuyuki Kato},
  {and} \bibinfo{person}{Yukitoshi Motome}.} \bibinfo{year}{2021}\natexlab{}.
\newblock \showarticletitle{Determinant-free fermionic wave function using
  feed-forward neural networks}.
\newblock \bibinfo{journal}{\emph{Physical Review Research}}
  \bibinfo{volume}{3}, \bibinfo{number}{4} (\bibinfo{year}{2021}),
  \bibinfo{pages}{043126}.
\newblock


\bibitem[\protect\citeauthoryear{Jia, Wang, Chen, Lu, Lin, Car, Weinan, and
  Zhang}{Jia et~al\mbox{.}}{2020}]%
        {jia2020pushing}
\bibfield{author}{\bibinfo{person}{Weile Jia}, \bibinfo{person}{Han Wang},
  \bibinfo{person}{Mohan Chen}, \bibinfo{person}{Denghui Lu},
  \bibinfo{person}{Lin Lin}, \bibinfo{person}{Roberto Car}, \bibinfo{person}{E
  Weinan}, {and} \bibinfo{person}{Linfeng Zhang}.}
  \bibinfo{year}{2020}\natexlab{}.
\newblock \showarticletitle{Pushing the limit of molecular dynamics with ab
  initio accuracy to 100 million atoms with machine learning}. In
  \bibinfo{booktitle}{\emph{SC20: International conference for high performance
  computing, networking, storage and analysis}}. IEEE, \bibinfo{pages}{1--14}.
\newblock


\bibitem[\protect\citeauthoryear{Jumper, Evans, Pritzel, Green, Figurnov,
  Ronneberger, Tunyasuvunakool, Bates, {\v{Z}}{\'\i}dek, Potapenko,
  et~al\mbox{.}}{Jumper et~al\mbox{.}}{2021}]%
        {jumper2021highly}
\bibfield{author}{\bibinfo{person}{John Jumper}, \bibinfo{person}{Richard
  Evans}, \bibinfo{person}{Alexander Pritzel}, \bibinfo{person}{Tim Green},
  \bibinfo{person}{Michael Figurnov}, \bibinfo{person}{Olaf Ronneberger},
  \bibinfo{person}{Kathryn Tunyasuvunakool}, \bibinfo{person}{Russ Bates},
  \bibinfo{person}{Augustin {\v{Z}}{\'\i}dek}, \bibinfo{person}{Anna
  Potapenko}, {et~al\mbox{.}}} \bibinfo{year}{2021}\natexlab{}.
\newblock \showarticletitle{Highly accurate protein structure prediction with
  AlphaFold}.
\newblock \bibinfo{journal}{\emph{Nature}} \bibinfo{volume}{596},
  \bibinfo{number}{7873} (\bibinfo{year}{2021}), \bibinfo{pages}{583--589}.
\newblock


\bibitem[\protect\citeauthoryear{Laughlin}{Laughlin}{1983}]%
        {Laugh1983}
\bibfield{author}{\bibinfo{person}{R.~B. Laughlin}.}
  \bibinfo{year}{1983}\natexlab{}.
\newblock \showarticletitle{Anomalous Quantum Hall Effect: An Incompressible
  Quantum Fluid with Fractionally Charged Excitations}.
\newblock \bibinfo{journal}{\emph{Phys. Rev. Lett.}}  \bibinfo{volume}{50}
  (\bibinfo{year}{1983}), \bibinfo{pages}{1395--1398}.
\newblock
Issue 18.


\bibitem[\protect\citeauthoryear{Li, Chen, Xiao, Wang, Jiang, Zhao, Lin, An,
  Liang, and He}{Li et~al\mbox{.}}{2022}]%
        {li2022bridging}
\bibfield{author}{\bibinfo{person}{Mingfan Li}, \bibinfo{person}{Junshi Chen},
  \bibinfo{person}{Qian Xiao}, \bibinfo{person}{Fei Wang},
  \bibinfo{person}{Qingcai Jiang}, \bibinfo{person}{Xuncheng Zhao},
  \bibinfo{person}{Rongfen Lin}, \bibinfo{person}{Hong An},
  \bibinfo{person}{Xiao Liang}, {and} \bibinfo{person}{Lixin He}.}
  \bibinfo{year}{2022}\natexlab{}.
\newblock \showarticletitle{Bridging the Gap between Deep Learning and
  Frustrated Quantum Spin System for Extreme-scale Simulations on New
  Generation of Sunway Supercomputer}.
\newblock \bibinfo{journal}{\emph{IEEE Transactions on Parallel and Distributed
  Systems}} (\bibinfo{year}{2022}).
\newblock


\bibitem[\protect\citeauthoryear{Li, Lin, Chen, Diaz, Xiao, Lin, Wang, Gao, and
  An}{Li et~al\mbox{.}}{2021}]%
        {li2021swflow}
\bibfield{author}{\bibinfo{person}{Mingfan Li}, \bibinfo{person}{Han Lin},
  \bibinfo{person}{Junshi Chen}, \bibinfo{person}{Jose~Monsalve Diaz},
  \bibinfo{person}{Qian Xiao}, \bibinfo{person}{Rongfen Lin},
  \bibinfo{person}{Fei Wang}, \bibinfo{person}{Guang~R Gao}, {and}
  \bibinfo{person}{Hong An}.} \bibinfo{year}{2021}\natexlab{}.
\newblock \showarticletitle{swFLOW: A large-scale distributed framework for
  deep learning on Sunway TaihuLight supercomputer}.
\newblock \bibinfo{journal}{\emph{Information Sciences}}  \bibinfo{volume}{570}
  (\bibinfo{year}{2021}), \bibinfo{pages}{831--847}.
\newblock


\bibitem[\protect\citeauthoryear{Liang, Dong, and He}{Liang
  et~al\mbox{.}}{2021}]%
        {liang2021hybrid}
\bibfield{author}{\bibinfo{person}{Xiao Liang}, \bibinfo{person}{Shao-Jun
  Dong}, {and} \bibinfo{person}{Lixin He}.} \bibinfo{year}{2021}\natexlab{}.
\newblock \showarticletitle{Hybrid convolutional neural network and projected
  entangled pair states wave functions for quantum many-particle states}.
\newblock \bibinfo{journal}{\emph{Physical Review B}} \bibinfo{volume}{103},
  \bibinfo{number}{3} (\bibinfo{year}{2021}), \bibinfo{pages}{035138}.
\newblock


\bibitem[\protect\citeauthoryear{Liang, Liu, Lin, Guo, Zhang, and He}{Liang
  et~al\mbox{.}}{2018}]%
        {liang2018solving}
\bibfield{author}{\bibinfo{person}{Xiao Liang}, \bibinfo{person}{Wen-Yuan Liu},
  \bibinfo{person}{Pei-Ze Lin}, \bibinfo{person}{Guang-Can Guo},
  \bibinfo{person}{Yong-Sheng Zhang}, {and} \bibinfo{person}{Lixin He}.}
  \bibinfo{year}{2018}\natexlab{}.
\newblock \showarticletitle{Solving frustrated quantum many-particle models
  with convolutional neural networks}.
\newblock \bibinfo{journal}{\emph{Physical Review B}} \bibinfo{volume}{98},
  \bibinfo{number}{10} (\bibinfo{year}{2018}), \bibinfo{pages}{104426}.
\newblock


\bibitem[\protect\citeauthoryear{Lin, Lin, Diaz, Li, An, and Gao}{Lin
  et~al\mbox{.}}{2019}]%
        {lin2019swflow}
\bibfield{author}{\bibinfo{person}{Han Lin}, \bibinfo{person}{Zeng Lin},
  \bibinfo{person}{Jose~Monsalve Diaz}, \bibinfo{person}{Mingfan Li},
  \bibinfo{person}{Hong An}, {and} \bibinfo{person}{Guang~R Gao}.}
  \bibinfo{year}{2019}\natexlab{}.
\newblock \showarticletitle{swFLOW: a dataflow deep learning framework on
  Sunway TaihuLight Supercomputer}. In \bibinfo{booktitle}{\emph{2019 IEEE 21st
  International Conference on High Performance Computing and Communications;
  IEEE 17th International Conference on Smart City; IEEE 5th International
  Conference on Data Science and Systems (HPCC/SmartCity/DSS)}}. IEEE,
  \bibinfo{pages}{2467--2475}.
\newblock


\bibitem[\protect\citeauthoryear{Liu, Gong, Li, Poilblanc, Chen, and Gu}{Liu
  et~al\mbox{.}}{2022}]%
        {LIU2022}
\bibfield{author}{\bibinfo{person}{Wen-Yuan Liu}, \bibinfo{person}{Shou-Shu
  Gong}, \bibinfo{person}{Yu-Bin Li}, \bibinfo{person}{Didier Poilblanc},
  \bibinfo{person}{Wei-Qiang Chen}, {and} \bibinfo{person}{Zheng-Cheng Gu}.}
  \bibinfo{year}{2022}\natexlab{}.
\newblock \showarticletitle{Gapless quantum spin liquid and global phase
  diagram of the spin-1/2 J1-J2 square antiferromagnetic Heisenberg model}.
\newblock \bibinfo{journal}{\emph{Science Bulletin}} (\bibinfo{year}{2022}).
\newblock
\showISSN{2095-9273}
\urldef\tempurl%
\url{https://doi.org/10.1016/j.scib.2022.03.010}
\showDOI{\tempurl}


\bibitem[\protect\citeauthoryear{Liu, Liu, Li, Fu, Yang, Song, Zhao, Wang,
  Peng, Chen, Guo, Huang, Wu, and Chen}{Liu et~al\mbox{.}}{2021}]%
        {GB2021}
\bibfield{author}{\bibinfo{person}{Yong~(Alexander) Liu},
  \bibinfo{person}{Xin~(Lucy) Liu}, \bibinfo{person}{Fang~(Nancy) Li},
  \bibinfo{person}{Haohuan Fu}, \bibinfo{person}{Yuling Yang},
  \bibinfo{person}{Jiawei Song}, \bibinfo{person}{Pengpeng Zhao},
  \bibinfo{person}{Zhen Wang}, \bibinfo{person}{Dajia Peng},
  \bibinfo{person}{Huarong Chen}, \bibinfo{person}{Chu Guo},
  \bibinfo{person}{Heliang Huang}, \bibinfo{person}{Wenzhao Wu}, {and}
  \bibinfo{person}{Dexun Chen}.} \bibinfo{year}{2021}\natexlab{}.
\newblock \showarticletitle{Closing the "Quantum Supremacy" Gap: Achieving
  Real-Time Simulation of a Random Quantum Circuit Using a New Sunway
  Supercomputer} \emph{(\bibinfo{series}{SC '21})}.
  \bibinfo{publisher}{Association for Computing Machinery},
  \bibinfo{address}{New York, NY, USA}, Article \bibinfo{articleno}{3},
  \bibinfo{numpages}{12}~pages.
\newblock
\showISBNx{9781450384421}
\urldef\tempurl%
\url{https://doi.org/10.1145/3458817.3487399}
\showDOI{\tempurl}


\bibitem[\protect\citeauthoryear{Luo and Clark}{Luo and Clark}{2019}]%
        {luo2019backflow}
\bibfield{author}{\bibinfo{person}{Di Luo} {and} \bibinfo{person}{Bryan~K
  Clark}.} \bibinfo{year}{2019}\natexlab{}.
\newblock \showarticletitle{Backflow transformations via neural networks for
  quantum many-body wave functions}.
\newblock \bibinfo{journal}{\emph{Physical review letters}}
  \bibinfo{volume}{122}, \bibinfo{number}{22} (\bibinfo{year}{2019}),
  \bibinfo{pages}{226401}.
\newblock


\bibitem[\protect\citeauthoryear{Moreno, Carleo, Georges, and Stokes}{Moreno
  et~al\mbox{.}}{2021}]%
        {moreno2021fermionic}
\bibfield{author}{\bibinfo{person}{Javier~Robledo Moreno},
  \bibinfo{person}{Giuseppe Carleo}, \bibinfo{person}{Antoine Georges}, {and}
  \bibinfo{person}{James Stokes}.} \bibinfo{year}{2021}\natexlab{}.
\newblock \showarticletitle{Fermionic Wave Functions from Neural-Network
  Constrained Hidden States}.
\newblock \bibinfo{journal}{\emph{arXiv preprint arXiv:2111.10420}}
  (\bibinfo{year}{2021}).
\newblock


\bibitem[\protect\citeauthoryear{Nomura and Imada}{Nomura and Imada}{2021}]%
        {nomura2021dirac}
\bibfield{author}{\bibinfo{person}{Yusuke Nomura} {and}
  \bibinfo{person}{Masatoshi Imada}.} \bibinfo{year}{2021}\natexlab{}.
\newblock \showarticletitle{Dirac-type nodal spin liquid revealed by refined
  quantum many-body solver using neural-network wave function, correlation
  ratio, and level spectroscopy}.
\newblock \bibinfo{journal}{\emph{Physical Review X}} \bibinfo{volume}{11},
  \bibinfo{number}{3} (\bibinfo{year}{2021}), \bibinfo{pages}{031034}.
\newblock


\bibitem[\protect\citeauthoryear{Savary and Balents}{Savary and
  Balents}{2017}]%
        {Savary2017}
\bibfield{author}{\bibinfo{person}{L. Savary} {and} \bibinfo{person}{L.
  Balents}.} \bibinfo{year}{2017}\natexlab{}.
\newblock \showarticletitle{Quantum spin liquids: a review}.
\newblock \bibinfo{journal}{\emph{Rep. Prog. Phys.}}  \bibinfo{volume}{80}
  (\bibinfo{year}{2017}), \bibinfo{pages}{016502}.
\newblock


\bibitem[\protect\citeauthoryear{Schollw\"ock}{Schollw\"ock}{2005}]%
        {DMRG}
\bibfield{author}{\bibinfo{person}{U. Schollw\"ock}.}
  \bibinfo{year}{2005}\natexlab{}.
\newblock \showarticletitle{The density-matrix renormalization group}.
\newblock \bibinfo{journal}{\emph{Rev. Mod. Phys.}}  \bibinfo{volume}{77}
  (\bibinfo{date}{Apr} \bibinfo{year}{2005}), \bibinfo{pages}{259--315}.
\newblock
Issue 1.
\urldef\tempurl%
\url{https://doi.org/10.1103/RevModPhys.77.259}
\showDOI{\tempurl}


\bibitem[\protect\citeauthoryear{Schollw\"ock}{Schollw\"ock}{2011}]%
        {Schollw11}
\bibfield{author}{\bibinfo{person}{U. Schollw\"ock}.}
  \bibinfo{year}{2011}\natexlab{}.
\newblock \showarticletitle{The density-matrix renormalization group in the age
  of matrix product states}.
\newblock \bibinfo{journal}{\emph{Annals of Physics}} \bibinfo{volume}{326},
  \bibinfo{number}{1} (\bibinfo{year}{2011}), \bibinfo{pages}{96 -- 192}.
\newblock


\bibitem[\protect\citeauthoryear{Silver, Huang, Maddison, Guez, Sifre, Van
  Den~Driessche, Schrittwieser, Antonoglou, Panneershelvam, Lanctot,
  et~al\mbox{.}}{Silver et~al\mbox{.}}{2016}]%
        {silver2016mastering}
\bibfield{author}{\bibinfo{person}{David Silver}, \bibinfo{person}{Aja Huang},
  \bibinfo{person}{Chris~J Maddison}, \bibinfo{person}{Arthur Guez},
  \bibinfo{person}{Laurent Sifre}, \bibinfo{person}{George Van Den~Driessche},
  \bibinfo{person}{Julian Schrittwieser}, \bibinfo{person}{Ioannis Antonoglou},
  \bibinfo{person}{Veda Panneershelvam}, \bibinfo{person}{Marc Lanctot},
  {et~al\mbox{.}}} \bibinfo{year}{2016}\natexlab{}.
\newblock \showarticletitle{Mastering the game of Go with deep neural networks
  and tree search}.
\newblock \bibinfo{journal}{\emph{nature}} \bibinfo{volume}{529},
  \bibinfo{number}{7587} (\bibinfo{year}{2016}), \bibinfo{pages}{484--489}.
\newblock


\bibitem[\protect\citeauthoryear{Silver, Schrittwieser, Simonyan, Antonoglou,
  Huang, Guez, Hubert, Baker, Lai, Bolton, et~al\mbox{.}}{Silver
  et~al\mbox{.}}{2017}]%
        {silver2017mastering}
\bibfield{author}{\bibinfo{person}{David Silver}, \bibinfo{person}{Julian
  Schrittwieser}, \bibinfo{person}{Karen Simonyan}, \bibinfo{person}{Ioannis
  Antonoglou}, \bibinfo{person}{Aja Huang}, \bibinfo{person}{Arthur Guez},
  \bibinfo{person}{Thomas Hubert}, \bibinfo{person}{Lucas Baker},
  \bibinfo{person}{Matthew Lai}, \bibinfo{person}{Adrian Bolton},
  {et~al\mbox{.}}} \bibinfo{year}{2017}\natexlab{}.
\newblock \showarticletitle{Mastering the game of go without human knowledge}.
\newblock \bibinfo{journal}{\emph{nature}} \bibinfo{volume}{550},
  \bibinfo{number}{7676} (\bibinfo{year}{2017}), \bibinfo{pages}{354--359}.
\newblock


\bibitem[\protect\citeauthoryear{Sorella}{Sorella}{1998}]%
        {SR}
\bibfield{author}{\bibinfo{person}{Sandro Sorella}.}
  \bibinfo{year}{1998}\natexlab{}.
\newblock \showarticletitle{Green Function Monte Carlo with Stochastic
  Reconfiguration}.
\newblock \bibinfo{journal}{\emph{Phys. Rev. Lett.}}  \bibinfo{volume}{80}
  (\bibinfo{date}{May} \bibinfo{year}{1998}), \bibinfo{pages}{4558--4561}.
\newblock
Issue 20.
\urldef\tempurl%
\url{https://doi.org/10.1103/PhysRevLett.80.4558}
\showDOI{\tempurl}


\bibitem[\protect\citeauthoryear{Steppke, K{\"u}chler, Lausberg, Lengyel,
  Steinke, Borth, L{\"u}hmann, Krellner, Nicklas, Geibel, Steglich, and
  Brando}{Steppke et~al\mbox{.}}{2013}]%
        {Steppke2013}
\bibfield{author}{\bibinfo{person}{A. Steppke}, \bibinfo{person}{R.
  K{\"u}chler}, \bibinfo{person}{S. Lausberg}, \bibinfo{person}{E. Lengyel},
  \bibinfo{person}{L. Steinke}, \bibinfo{person}{R. Borth}, \bibinfo{person}{T.
  L{\"u}hmann}, \bibinfo{person}{C. Krellner}, \bibinfo{person}{M. Nicklas},
  \bibinfo{person}{C. Geibel}, \bibinfo{person}{F. Steglich}, {and}
  \bibinfo{person}{M. Brando}.} \bibinfo{year}{2013}\natexlab{}.
\newblock \showarticletitle{Ferromagnetic Quantum Critical Point in the
  Heavy-Fermion Metal YbNi4(P1-xAsx)2}.
\newblock \bibinfo{journal}{\emph{Science}} \bibinfo{volume}{339},
  \bibinfo{number}{6122} (\bibinfo{year}{2013}), \bibinfo{pages}{933--936}.
\newblock


\bibitem[\protect\citeauthoryear{Stokes, Moreno, Pnevmatikakis, and
  Carleo}{Stokes et~al\mbox{.}}{2020}]%
        {stokes2020phases}
\bibfield{author}{\bibinfo{person}{James Stokes},
  \bibinfo{person}{Javier~Robledo Moreno}, \bibinfo{person}{Eftychios~A
  Pnevmatikakis}, {and} \bibinfo{person}{Giuseppe Carleo}.}
  \bibinfo{year}{2020}\natexlab{}.
\newblock \showarticletitle{Phases of two-dimensional spinless lattice fermions
  with first-quantized deep neural-network quantum states}.
\newblock \bibinfo{journal}{\emph{Physical Review B}} \bibinfo{volume}{102},
  \bibinfo{number}{20} (\bibinfo{year}{2020}), \bibinfo{pages}{205122}.
\newblock


\bibitem[\protect\citeauthoryear{Stormer}{Stormer}{1999}]%
        {Stormer1999}
\bibfield{author}{\bibinfo{person}{H.~L. Stormer}.}
  \bibinfo{year}{1999}\natexlab{}.
\newblock \showarticletitle{Nobel Lecture: The fractional quantum Hall effect}.
\newblock \bibinfo{journal}{\emph{Rev. Mod. Phys.}}  \bibinfo{volume}{71}
  (\bibinfo{year}{1999}), \bibinfo{pages}{875--889}.
\newblock
Issue 4.


\bibitem[\protect\citeauthoryear{Szab{\'o} and Castelnovo}{Szab{\'o} and
  Castelnovo}{2020}]%
        {szabo2020neural}
\bibfield{author}{\bibinfo{person}{Attila Szab{\'o}} {and}
  \bibinfo{person}{Claudio Castelnovo}.} \bibinfo{year}{2020}\natexlab{}.
\newblock \showarticletitle{Neural network wave functions and the sign
  problem}.
\newblock \bibinfo{journal}{\emph{Physical Review Research}}
  \bibinfo{volume}{2}, \bibinfo{number}{3} (\bibinfo{year}{2020}),
  \bibinfo{pages}{033075}.
\newblock


\bibitem[\protect\citeauthoryear{Troyer and Wiese}{Troyer and Wiese}{2005}]%
        {Troyer05}
\bibfield{author}{\bibinfo{person}{M. Troyer} {and} \bibinfo{person}{U.-J.
  Wiese}.} \bibinfo{year}{2005}\natexlab{}.
\newblock \showarticletitle{Computational Complexity and Fundamental
  Limitations to Fermionic Quantum Monte Carlo Simulations}.
\newblock \bibinfo{journal}{\emph{Phys. Rev. Lett.}}  \bibinfo{volume}{94}
  (\bibinfo{year}{2005}), \bibinfo{pages}{170201}.
\newblock
Issue 17.


\bibitem[\protect\citeauthoryear{Tsui, Stormer, and Gossard}{Tsui
  et~al\mbox{.}}{1982}]%
        {Tsui1982}
\bibfield{author}{\bibinfo{person}{D.~C. Tsui}, \bibinfo{person}{H.~L.
  Stormer}, {and} \bibinfo{person}{A.~C. Gossard}.}
  \bibinfo{year}{1982}\natexlab{}.
\newblock \showarticletitle{Two-Dimensional Magnetotransport in the Extreme
  Quantum Limit}.
\newblock \bibinfo{journal}{\emph{Phys. Rev. Lett.}}  \bibinfo{volume}{48}
  (\bibinfo{year}{1982}), \bibinfo{pages}{1559--1562}.
\newblock
Issue 22.


\bibitem[\protect\citeauthoryear{Verstraete, Murg, and Cirac}{Verstraete
  et~al\mbox{.}}{2008}]%
        {PEPS}
\bibfield{author}{\bibinfo{person}{F. Verstraete}, \bibinfo{person}{V. Murg},
  {and} \bibinfo{person}{J.I. Cirac}.} \bibinfo{year}{2008}\natexlab{}.
\newblock \showarticletitle{Matrix product states, projected entangled pair
  states, and variational renormalization group methods for quantum spin
  systems}.
\newblock \bibinfo{journal}{\emph{Advances in Physics}} \bibinfo{volume}{57},
  \bibinfo{number}{2} (\bibinfo{year}{2008}), \bibinfo{pages}{143--224}.
\newblock
\urldef\tempurl%
\url{https://doi.org/10.1080/14789940801912366}
\showDOI{\tempurl}
\showeprint{https://doi.org/10.1080/14789940801912366}


\bibitem[\protect\citeauthoryear{Wang and Sandvik}{Wang and Sandvik}{2018}]%
        {dmrgj1j2}
\bibfield{author}{\bibinfo{person}{Ling Wang} {and} \bibinfo{person}{Anders~W.
  Sandvik}.} \bibinfo{year}{2018}\natexlab{}.
\newblock \showarticletitle{Critical Level Crossings and Gapless Spin Liquid in
  the Square-Lattice Spin-$1/2$ ${J}_{1}\ensuremath{-}{J}_{2}$ Heisenberg
  Antiferromagnet}.
\newblock \bibinfo{journal}{\emph{Phys. Rev. Lett.}}  \bibinfo{volume}{121}
  (\bibinfo{date}{Sep} \bibinfo{year}{2018}), \bibinfo{pages}{107202}.
\newblock
Issue 10.
\urldef\tempurl%
\url{https://doi.org/10.1103/PhysRevLett.121.107202}
\showDOI{\tempurl}


\bibitem[\protect\citeauthoryear{Wen}{Wen}{1992}]%
        {wen1992theory}
\bibfield{author}{\bibinfo{person}{Xiao-Gang Wen}.}
  \bibinfo{year}{1992}\natexlab{}.
\newblock \showarticletitle{Theory of the edge states in fractional quantum
  Hall effects}.
\newblock \bibinfo{journal}{\emph{International journal of modern physics B}}
  \bibinfo{volume}{6}, \bibinfo{number}{10} (\bibinfo{year}{1992}),
  \bibinfo{pages}{1711--1762}.
\newblock


\bibitem[\protect\citeauthoryear{White}{White}{1992}]%
        {white92}
\bibfield{author}{\bibinfo{person}{S.~R. White}.}
  \bibinfo{year}{1992}\natexlab{}.
\newblock \showarticletitle{Density matrix formulation for quantum
  renormalization groups}.
\newblock \bibinfo{journal}{\emph{Phys. Rev. Lett.}}  \bibinfo{volume}{69}
  (\bibinfo{year}{1992}), \bibinfo{pages}{2863--2866}.
\newblock
Issue 19.


\bibitem[\protect\citeauthoryear{Zhou, Lee, Imada, Trivedi, Phillips, Kee,
  T{\"o}rm{\"a}, and Eremets}{Zhou et~al\mbox{.}}{2021}]%
        {zhou2021high}
\bibfield{author}{\bibinfo{person}{Xingjiang Zhou}, \bibinfo{person}{Wei-Sheng
  Lee}, \bibinfo{person}{Masatoshi Imada}, \bibinfo{person}{Nandini Trivedi},
  \bibinfo{person}{Philip Phillips}, \bibinfo{person}{Hae-Young Kee},
  \bibinfo{person}{P{\"a}ivi T{\"o}rm{\"a}}, {and} \bibinfo{person}{Mikhail
  Eremets}.} \bibinfo{year}{2021}\natexlab{}.
\newblock \showarticletitle{High-temperature superconductivity}.
\newblock \bibinfo{journal}{\emph{Nature Reviews Physics}} \bibinfo{volume}{3},
  \bibinfo{number}{7} (\bibinfo{year}{2021}), \bibinfo{pages}{462--465}.
\newblock


\end{thebibliography}
\end{document}